# *Data science to investigate temperature profiles of large networks of food refrigeration systems*

Corneliu Arsene[1,2]

Email : corneliuarsene@gmax.com, arsenecorneliu@tutanota.de

University of Manchester, Manchester, UK

## I. Introduction

The electrical generation and transmission infrastructures of many countries are under increased pressure [1]. This partially reflects the move towards low carbon economies and the increased reliance on renewable power generation systems [2]. There has been a reduction in the use of traditional fossil fuel generation systems, which provide a stable base load, and this has been replaced with more unpredictable renewable generation [3]. As a consequence, the available load on the grid is becoming more unstable. To cope with this variability, the UK National Grid has placed emphasis on the investigation of various technical mechanisms (e.g. implementation of smart grids, energy storage technologies, auxiliary power sources, Demand Side Response (DSR)) [4-6], which may be able to prevent critical situations, when the grid may become sometimes unstable. The successful implementation of these mechanisms may require large numbers of electrical consumers (e.g. HVAC systems, food refrigeration systems) for example to make additional investments in energy storage technologies (i.e. food refrigeration systems) or to integrate their electrical demand from industrial processes into the National Grid (i.e. HVAC systems). However, for food refrigeration systems, during these critical situations, even if the thermal inertia within refrigeration systems may maintain effective performance of the device for a short period of time (e.g. under 1 minute) when the electrical input load into the system is reduced, this still carries the paramount risk of food safety even for very short periods of time (e.g. 1 under minute). Therefore before considering any future actions (e.g.

---

[1]Present affiliation
[2]Work implemented at UoL, UK.

investing in energy storage technologies) to prevent the critical situations when grid becomes unstable, it is also needed to understand during the normal use how the temperature profiles evolve along the time inside these massive networks of food refrigeration systems during either shorter (i.e. minutes) or longer periods of time (i.e. days, months) and this paper presents this.   It will be made a difference between the refrigeration cases termed as Low Temperature (LT) refrigeration cases with temperature inside them around -18 degrees  Celsius and the High Temperature (HT) refrigeration cases with temperature inside them around 5 degrees  Celsius.   The paper is structured as follows: the main gist of the paper is in section II where it is presented data science for massive networks of LT and HT food refrigeration systems.  Section III makes a short overview on how Big data and Internet of Things (IoTs) may play a role in the implementation of controlling and monitoring systems of massive networks of LT and HT food refrigeration systems and then paper ends with some brief conclusions (section IV).

## II. DATA SCIENCE FOR INVESTIGATING TEMPERATURE PROFILES OF MASSIVE NETWORKS OF LT AND HT FOOD REFRIGERATION SYSTEMS

Modelling and simulation of large, complex and nonlinear systems poses significant challenges [7-15] even when simulated data is used.   Therefore employing real data in this process can only alleviate this task.  In this section, it will be used data science in order to investigate the temperature profiles of massive networks of LT and HT food refrigeration systems and without relying on simulated data.  This investigation can help further to understand how to monitor and control such large, complex and nonlinear systems as the massive networks of food refrigeration systems.

First, a brief description is made of a single refrigeration system (i.e. case) [16] typically used in the food retailer sector.  Figure 1 shows the four main components of a refrigeration system: expansion valve, evaporator, gas compressor, condenser.  Each component fulfils a function, which forms a four-step refrigeration cycle. The liquid refrigerant goes through the expansion valve where pressure is reduced as well as the temperature drops down. The cooled refrigerant liquid together with any vapour (i.e. gas) then goes to evaporator where the liquid refrigerant takes away the heat by boiling to vapour (i.e. gas) which in turns goes to the Compressor. In the Compressor the pressure is increased so that the refrigerant to compress and become a hot temperature higher pressure gas. This hot high pressure gas goes to the Condenser where it liquefies into a high pressure liquid, which may go back to the Expansion valve forming the four step refrigeration cycle.

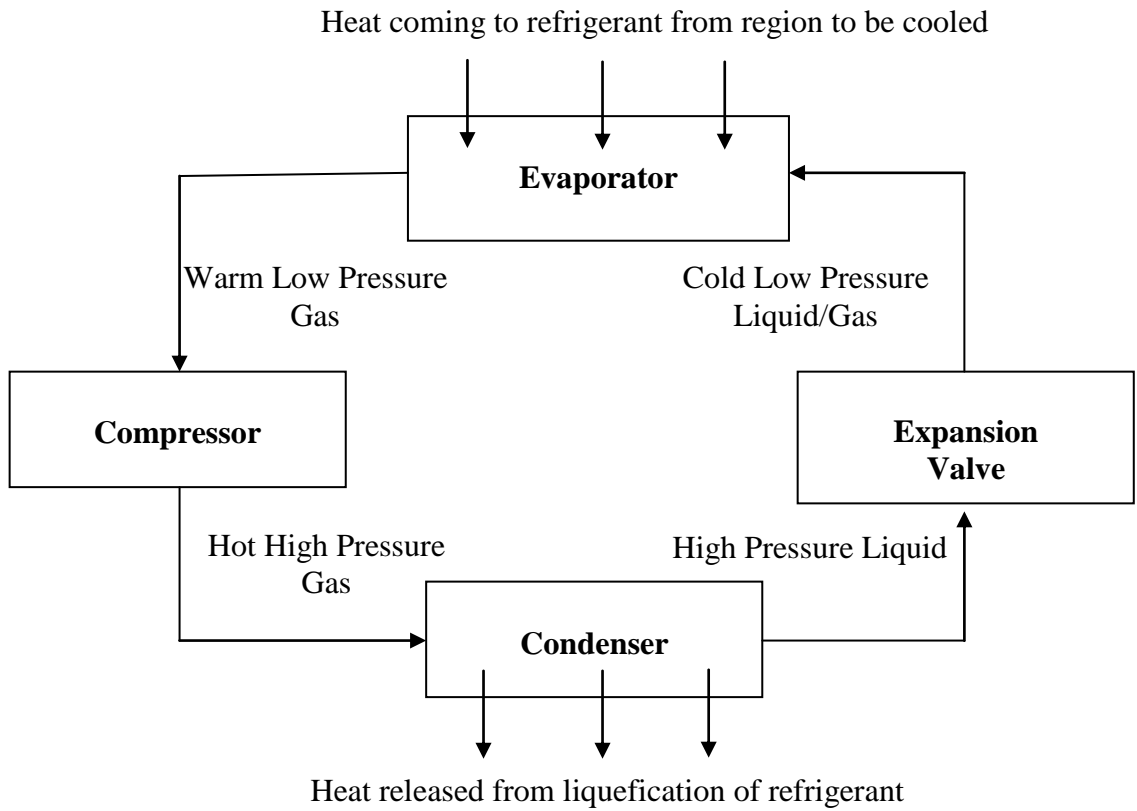

Figure 1. Four components which form a refrigeration system.

*A. Single time snapshot*

Figure 2 shows the typical temperature distributions of an aggregated refrigeration network belonging to an important food retailer from UK at any instant in time. The figure contains data for several cases types, which contain a range of fresh foods.

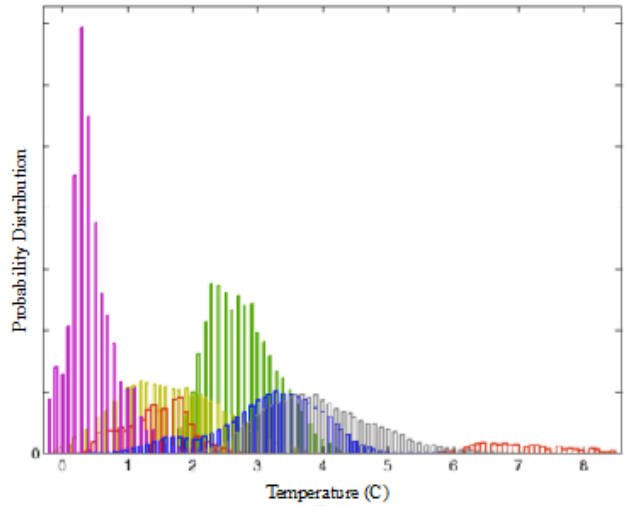

Figure 2. The distribution of refrigeration case temperatures taken at an instant in time for 10000s of refrigeration cases in the main retail estate: 7 different types of food are classifed with data shown by different colours.

Figure 2 shows that large food retailers have a significant number of refrigeration control regimes for different products. This reflects both the temperature required to maintain the food properly that maximises product shelf life, but also ensures food safety, and fresh products can for example be maintained at slightly higher temperatures than dairy. For different refrigeration cases the set point control band can be wide, and the actual achieved temperature for the population of refrigerators within the estate varies within that control band (shown in Figure 2). Within Figure 2 it is possible to see that certain classes of products have actual set point within a narrow control band. For example, note the spike on the left where the product temperature is 0.4°C but the population of cases are +/- 1°C, compared to the foods cases on the right of figure where mean control points are 3.5°C but the population sits within a window +/- 3°C of the mean.

The temperatures profiles shown in Figure 2 are the Computed Product Temperatures (CPTs) which are meant to represent the temperature of the food inside a refrigeration case and is calculated based on the air-on ($T_{air-on}$) and the air-off ($T_{air-off}$) temperatures from inside the refrigeration case/system [17, 18]: the "air-on" sensor temperature indicates the temperature of air going towards the evaporator of the refrigeration case, while the "air-off" sensor indicates the temperature of air leaving the evaporator of the refrigeration case. All these measures (i.e. CPT temperature, air-off temperature, air-on temperature) could be provided either off-line in various database formats (i.e. text files, binary files, SQL databases, etc) by the commercial and grocery chain of hyper and supermarkets or in real time. Actually, in Figure 3 it can be seen how a typical refrigeration case may look in real life and the typical location of the air-off and air-on sensors while the other measures are as follows: $T_{amb}$ is the ambient indoor temperature, $T_{pr}$ is the product temperature, $T_{air}$ is the air temperature inside the case, $T_e$ is the evaporation temperature, $\dot{H}_e$ is the cooling capacity, $\dot{H}_{pr}$ is the heat transfer between the product and the air inside the case, $\dot{H}_{amb}$ is the heat load from the environment, T$_{shelf}$ is the shelf temperature (i.e. T$_{shelf}$= 0.6 T$_{air-off}$ + 0.4 T$_{air-on}$). In this context the CPT is calculated based on T$_{shelf}$ at k-th minute and the previous 29 CPT$_{k-1}$ calculations (i.e. $CPT_k = \frac{1}{30}\left(T_{shelf}^k + \sum_{i=1}^{29} CPT_{k-1}\right)$) [17, 18].

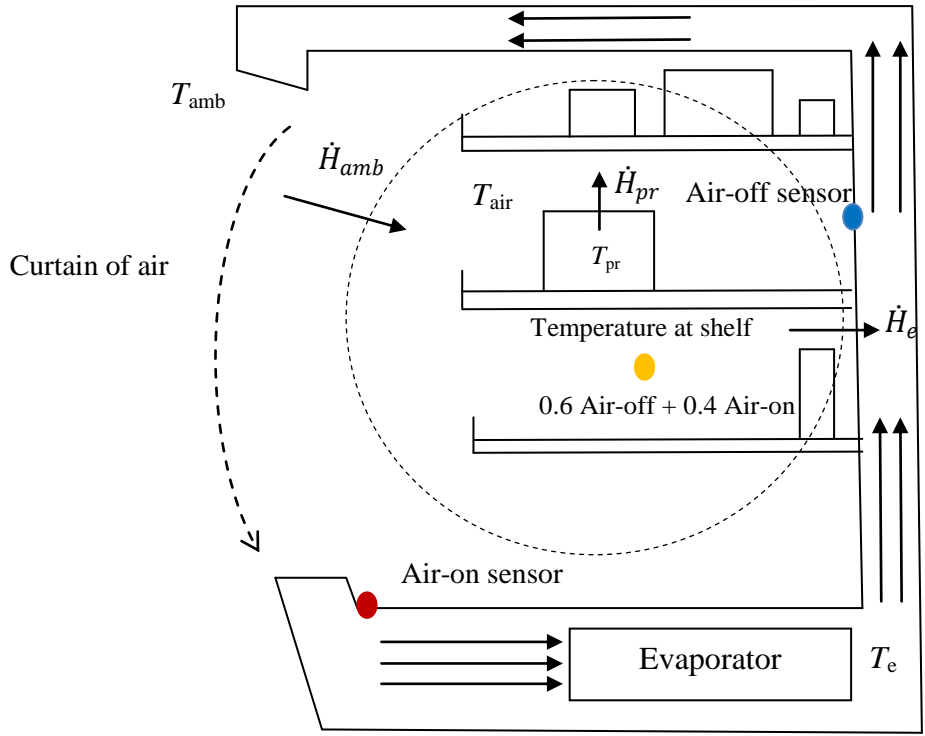

Figure 3. Drawing of a typical refrigeration case located in food retailer shops.

Figure 4 shows the number of cases and the temperature distribution (i.e. CPT) for a different set of 63722 refrigeration cases at one time snapshot.

It can be noticed clearly that the two clusters formed around -20 degrees Celsius for a first cluster of Low Temperature (LT) cases and the second cluster formed around 2.99 degrees Celsius for High Temperature (HT) cases. Table 1 shows the bins formed by clustering the data and using 100 bins as well as the position (i.e. CPT temperature) of the bin centers. The interest in this paper is to study the differences or the similarities in the temperature profiles along the time of these HT and LT refrigeration cases.

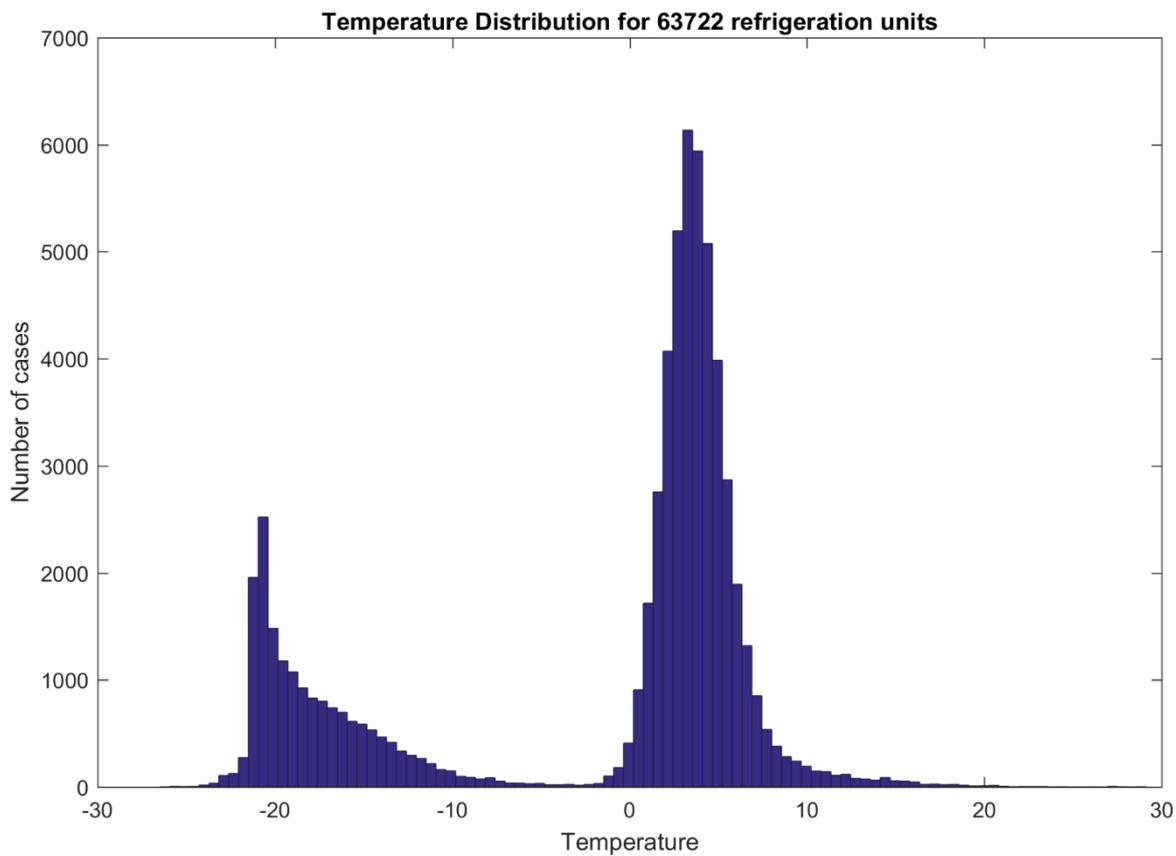

Figure 4. Number of cases and CPT temperature distribution for 63722 refrigeration cases at one moment of time.

| N | Temp | N | Temp | N | Temp | N | Temp | N | Temp |
|---|---|---|---|---|---|---|---|---|---|
| 1 | -26.23 | 587 | -15.09 | 20 | -3.96 | 852 | 7.17 | 23 | 18.30 |
| 6 | -25.67 | 534 | -14.53 | 24 | -3.40 | 538 | 7.72 | 16 | 18.86 |
| 4 | -25.11 | 466 | -13.98 | 16 | -2.84 | 380 | 8.28 | 10 | 19.41 |
| 6 | -24.55 | 417 | -13.42 | 23 | -2.92 | 283 | 8.84 | 12 | 19.97 |
| 17 | -24.00 | 335 | -12.86 | 33 | -1.73 | 241 | 9.39 | 15 | 20.53 |
| 35 | -23.44 | 296 | -12.31 | 103 | -1.17 | 193 | 9.95 | 7 | 21.08 |
| 107 | -22.88 | 266 | -11.75 | 181 | -0.62 | 149 | 10.51 | 3 | 21.64 |
| 126 | -22.33 | 218 | -11.19 | 410 | -0.06 | 143 | 11.06 | 4 | 22.20 |
| 275 | -21.77 | 162 | -10.64 | 907 | 0.49 | 109 | 11.62 | 4 | 22.75 |
| 1957 | -21.21 | 151 | -10.08 | 1716 | 1.04 | 117 | 12.18 | 4 | 23.31 |
| 2522 | -20.66 | 99 | -9.52 | 2757 | 1.60 | 81 | 12.73 | 1 | 23.87 |
| 1480 | -20.10 | 89 | -8.97 | 4071 | 2.16 | 74 | 13.29 | 2 | 24.42 |
| 1179 | -19.54 | 75 | -8.41 | 5195 | 2.71 | 63 | 13.85 | 0 | 24.98 |
| 1074 | -18.99 | 86 | -7.85 | 6135 | 3.27 | 88 | 14.40 | 0 | 25.54 |
| 925 | -18.43 | 55 | -7.30 | 5941 | 3.83 | 58 | 14.96 | 1 | 26.09 |
| 830 | -17.87 | 37 | -6.74 | 5077 | 4.38 | 56 | 15.52 | 0 | 27.65 |
| 802 | -17.32 | 36 | -6.18 | 3985 | 4.94 | 46 | 16.07 | 6 | 27.21 |
| 739 | -16.76 | 31 | -5.63 | 2869 | 5.50 | 24 | 16.63 | 2 | 27.76 |
| 698 | -16.20 | 34 | -5.07 | 1894 | 6.05 | 27 | 17.19 | 0 | 28.32 |
| 613 | -15.65 | 20 | -4.51 | 1320 | 6.61 | 21 | 17.74 | 2 | 28.88 |

Table 1. Clustering the data in 100 bins and the mean CPT corresponding to each bin.

From the entire estate of 63722 cases, there are selected the cases which are labelled as frozen food and identified as FFW, FFF, FFGD, LT, FFGDND, FFWND, FFFND: they will all be generally termed as LT refrigeration cases/cases. This resulted in a subset of 17585 in which the CPT should be very low below -5 degrees Celsius. Again, similar to Figure 4, the temperature distribution of these 17585 cases is shown in Figure 5. There are used 50 bins.

From the entire subset of 17585 LT cases, about 272 cases had the CPT higher than -5 degrees Celsius which is due either to the case being in the defrost or either to errors in measuring/acquiring the CPT (Figure 6).

From the entire estate of 63722 cases, there are selected the cases (i.e. 46137 cases) which are labelled as higher temperature cases and identified in various ways such as 'MFM', 'HT', 'AMB', 'DAI', 'PDR', 'FAV', 'BWS', 'BUC', 'DEL', 'DAIND', 'DBK', 'FMSO', 'AMBND', 'FAVND', 'BUCND', 'DELND', 'FMSOND', 'DBKND': they will all be generally termed as HT refrigeration cases/cases. This resulted in a subset of 46137 cases in which the CPT should be higher than 0 degrees Celsius. Again, similar to Figure 4, the temperature distribution of these 46137 cases is shown in Figure 7. There are used again 50 bins.

From the entire subset of 46137 HT cases, about 45 cases had the CPT lower than -5 degrees Celsius which is due either to the lowering of the case temperature or either to errors in measuring/acquiring the CPT (Figure 8).

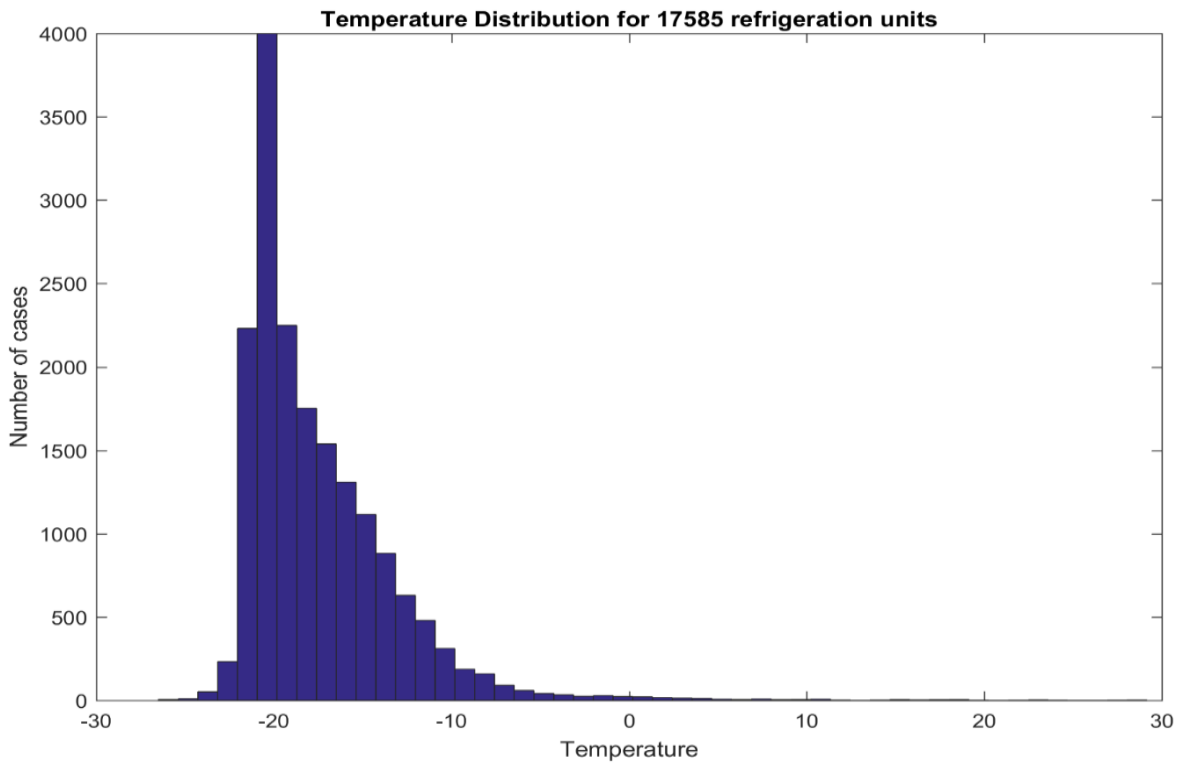

Figure 5. CPT temperature distribution for 17585 cases at one time snapshot and identified as lower temperature cases (FFW, FFF, FFGD, LT, FFGDND, FFWND, FFFND).

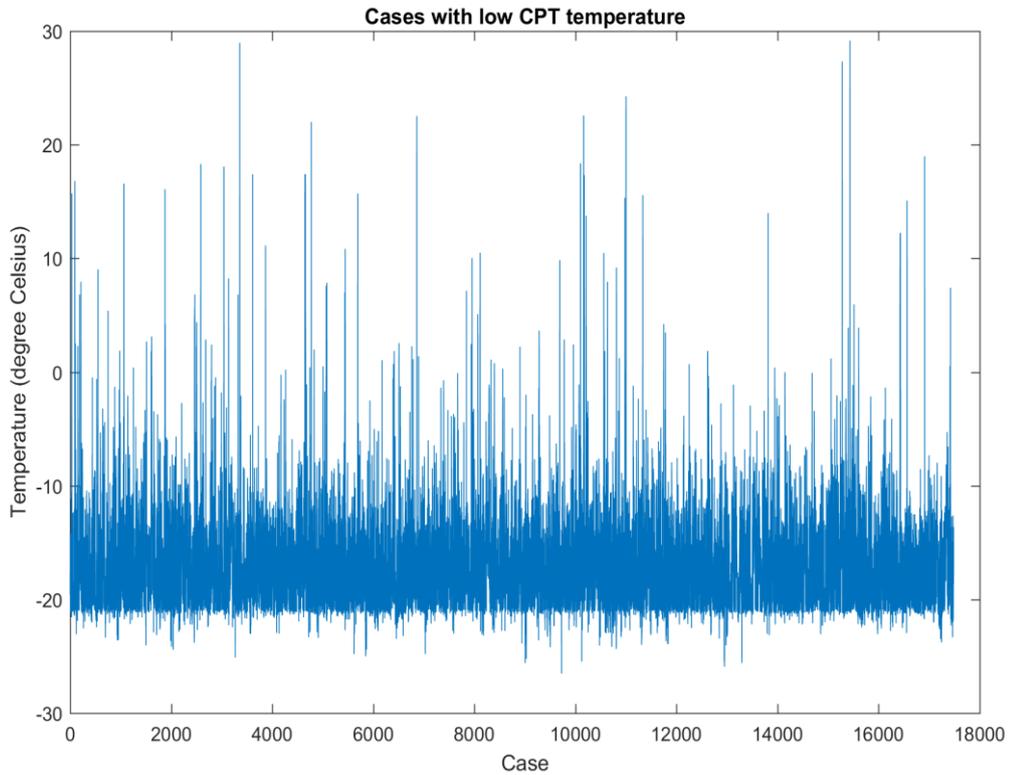

Figure 6. CPT temperature distribution for 17585 cases at one time snapshot, identified as lower temperature cases and with 272 cases which have CPT higher than -5 degrees Celsius (i.e. the defrost situation or either errors in measuring/acquiring the CPT).

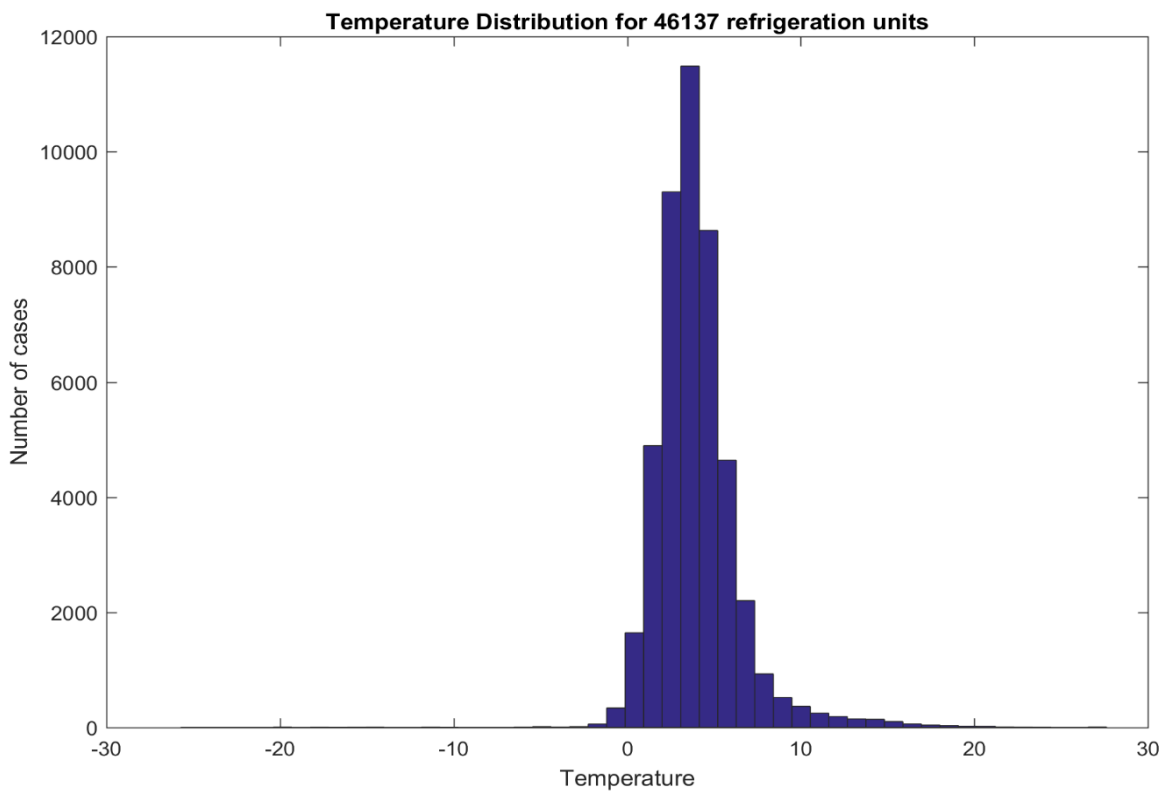

Figure 7. CPT temperature distribution for 46137 cases at one time snapshot and identified as higher temperature cases.

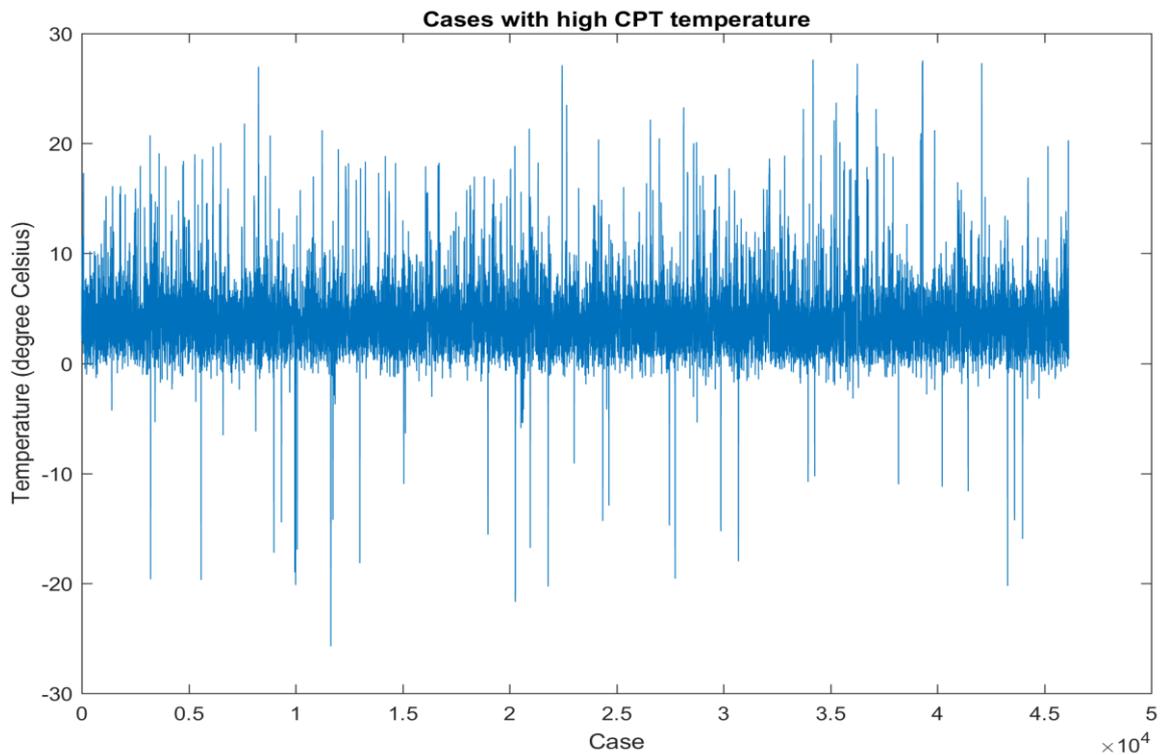

Figure 8. CPT temperature distribution for 46137 cases at one time snapshot, identified as higher temperature cases and with 45 cases which have CPT lower than -5 degrees Celsius.

Furthermore, Figure 9 shows the number of cases and the temperature (CPT) distribution for 46137 cases at one time snapshot and identified as higher temperature cases together with a probability distribution fit based on a non-parametric kernel-smoothing distribution in Matlab software. Similarly, Figure 10 shows the number of cases and the temperature (CPT) distribution for 17585 cases at one time snapshot and identified as lower temperature cases together again with a probability distribution fit based on a non-parametric kernel-smoothing distribution in Matlab software.

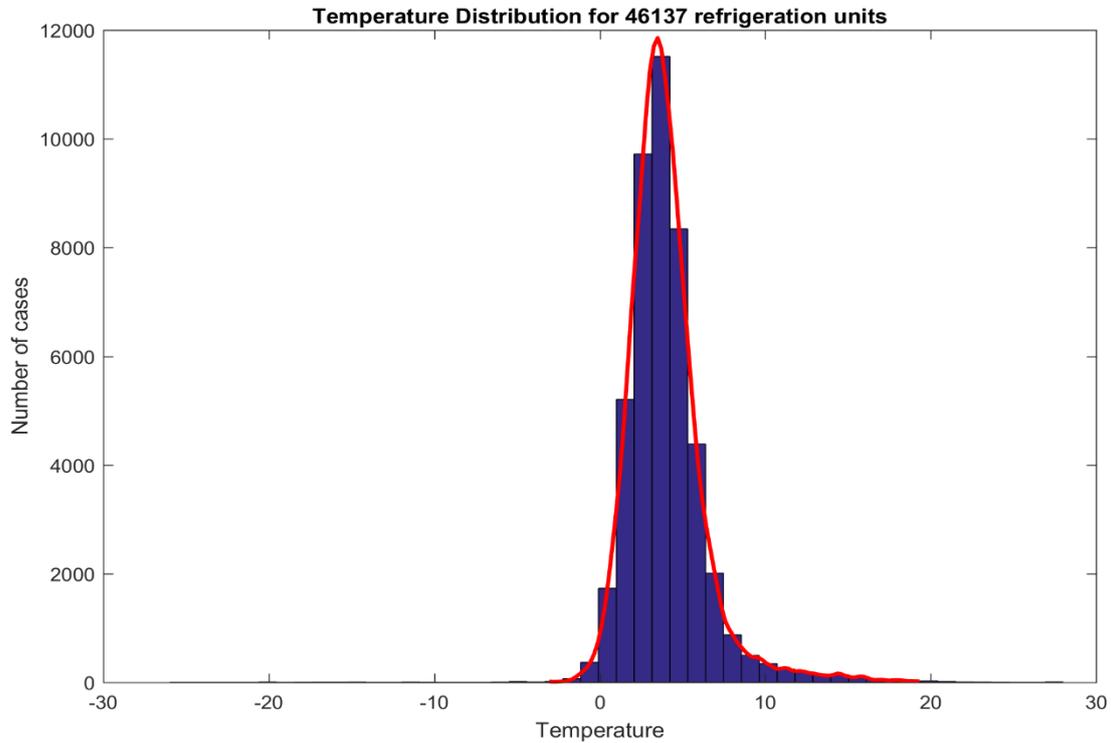

Figure 9. Number of cases and CPT temperature distribution for 46137 cases at one time snapshot and identified as higher temperature cases together with a probability distribution fit based on a non-parametric kernel-smoothing distribution.

Figure 11 and Figure 12 show the number of cases together with probability distribution fits for the different types of LT 17585 cases FFF (2320 cases), FFWND (37 cases), FFGDND (56 cases), FFGD (6500 cases), FFW (8624 cases), LT (39 cases) and FFFND (9 cases) shown per case type.

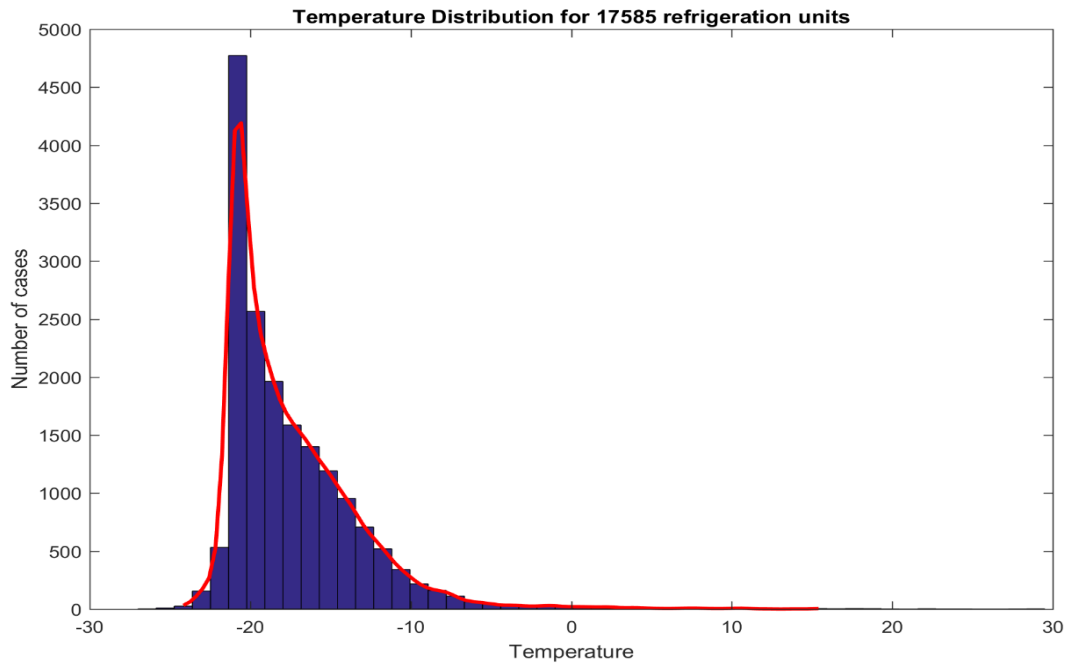

Figure 10. Number of cases and CPT temperature distribution for 17585 cases at one time snapshot and identified as lower temperature cases together with a probability distribution fit based on a non-parametric kernel-smoothing distribution.

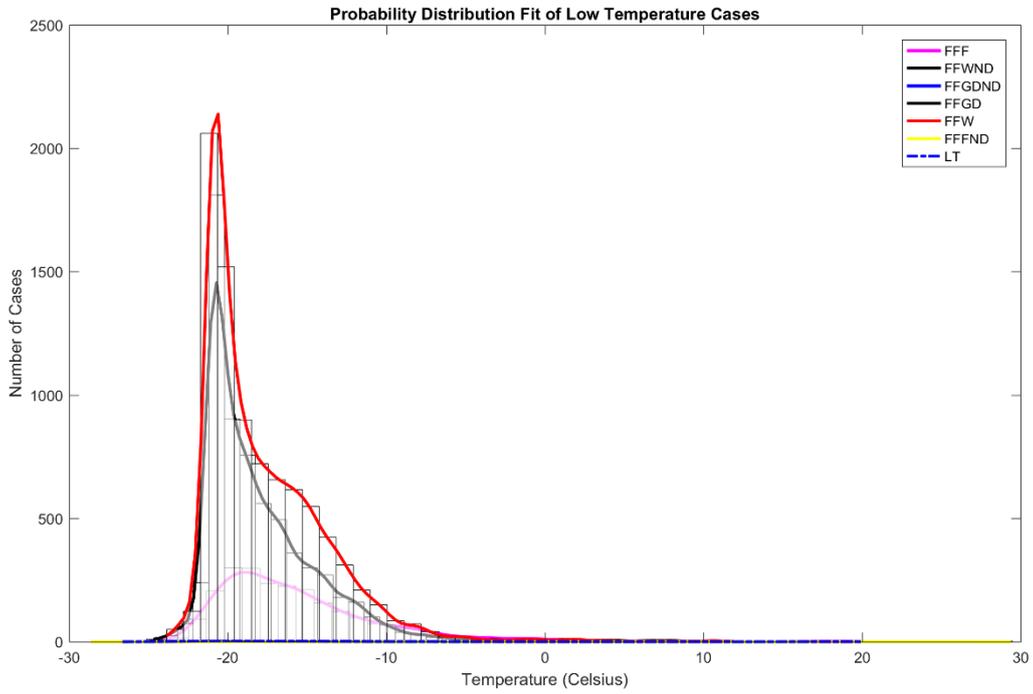

Figure 11. Number of cases together with probability distributions of different types of LT 17585 cases (FFF, FFWND, FFGDND, FFGD, FFW, FFFND, LT) function of CPT.

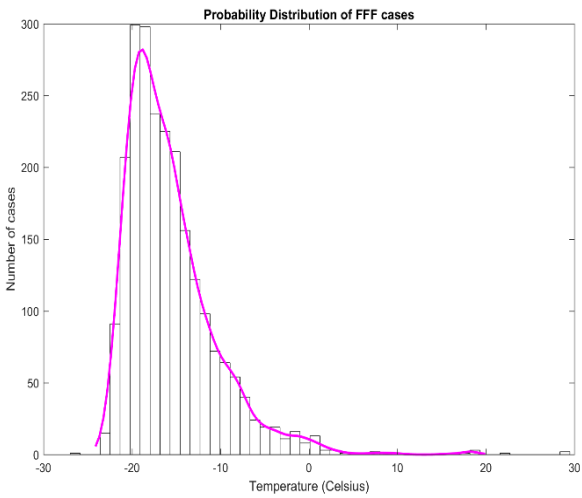
a) FFF

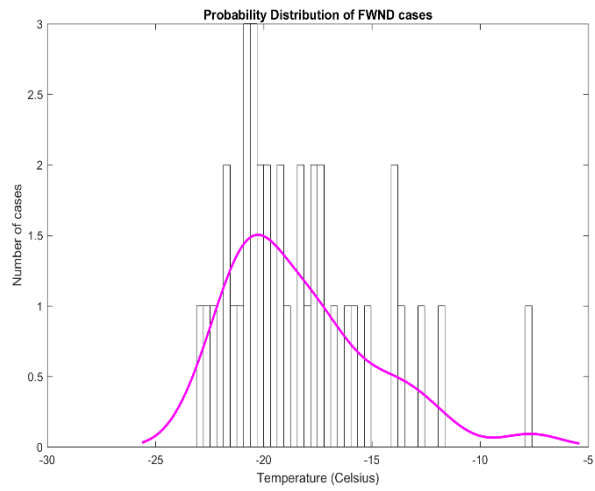
b) FWND

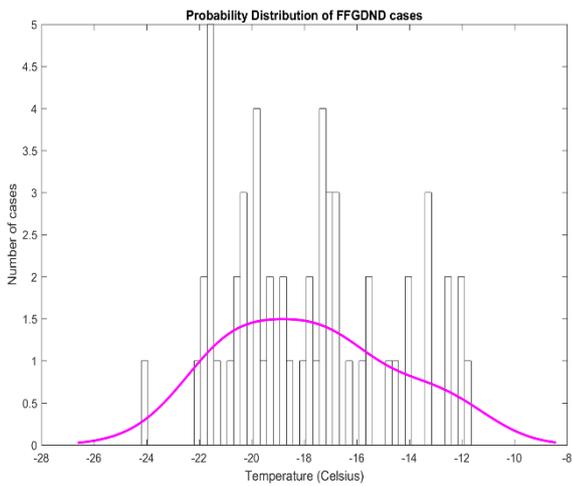
c) FFGDND

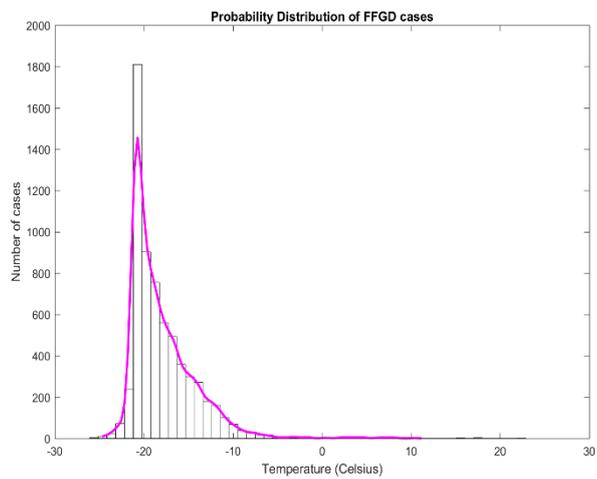
d) FFGD

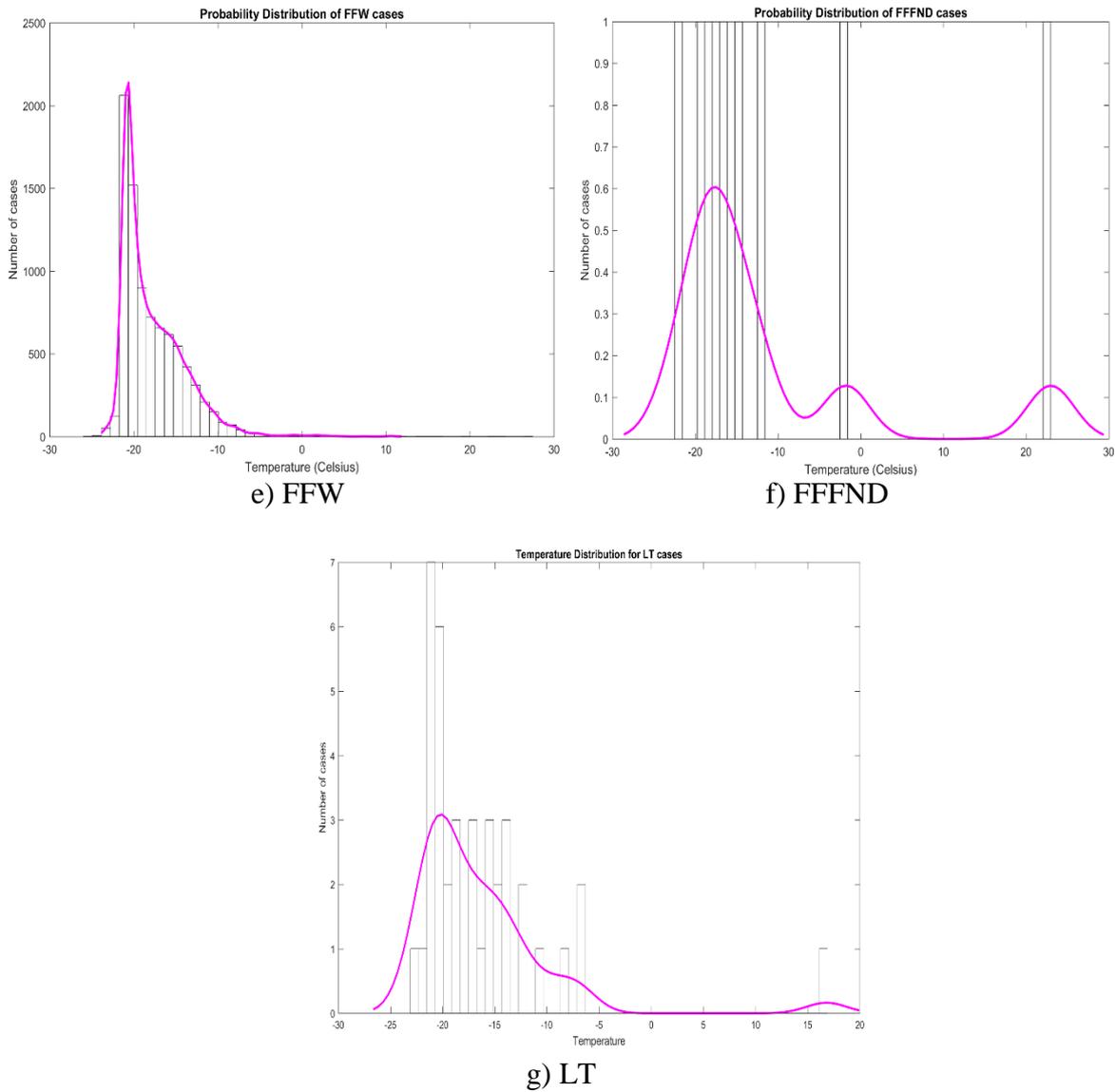

Figure 12. Number of cases together with probability distributions of LT 17585 cases function of CPT temperature: FFF (2320 cases), FFWND (37 cases), FFGDND (56 cases), FFGD (6500 cases), FFW (8624 cases), FFFND (9 cases), LT (39 cases).

Figures 13, 14 and 15 show the number of cases and probability distribution fits for the various HT cases totalizing 46137 cases and listed herein: 'MFM' (12131 cases), 'HT' (335 cases), 'AMB' (240 cases), 'DAI' (18226 cases), 'PDR' (2622 cases), 'FAV' (8683 cases), 'BWS' (53 cases), 'BUC' (1005 cases), 'DEL' (1917 cases), 'DAIND' (62 cases), 'DBK' (709 cases), 'FMSO' (109 cases), 'AMBND' (6 cases), 'FAVND' (19 cases), 'BUCND' (3 cases), 'DELND' (5 cases), 'FMSOND' (2 cases), 'DBKND' (2 cases). Figure 16 shows the number of cases and all the 18 probability distributions for the HT cases.

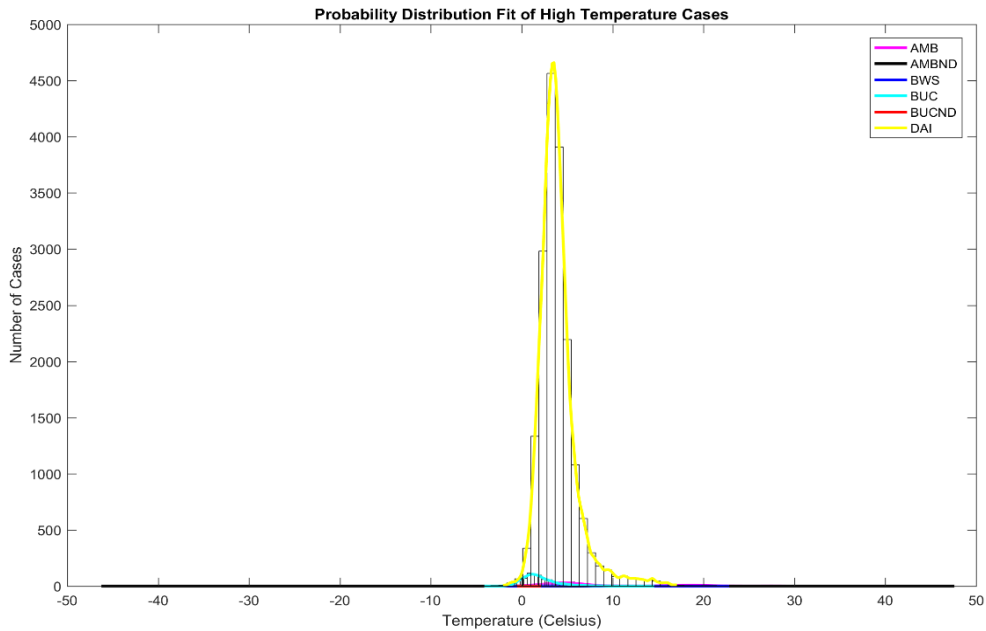

Figure 13. Number of cases and probability distributions of the following HT cases function of CPT temperature: AMB (240 cases), AMBND (6 cases), BWS (53 cases), BUC (1005 cases), BUCND (3 cases), DAI (18226 cases).

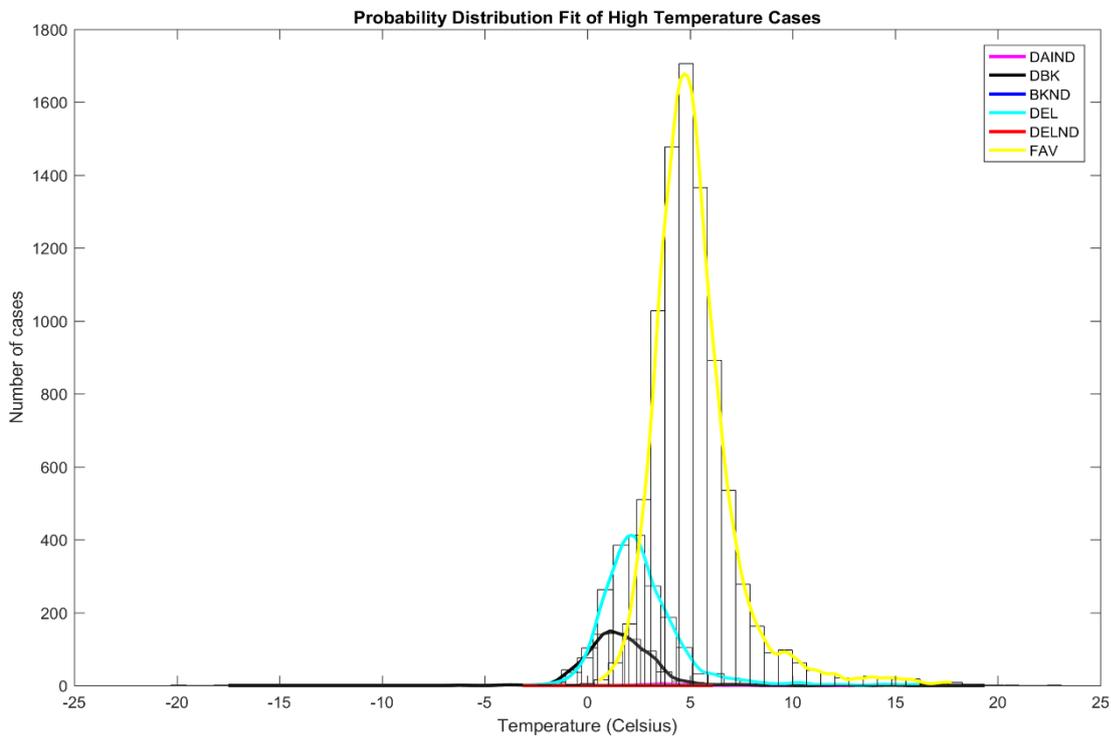

Figure 14. Number of cases and CPT probability distributions of the following types of HT cases: DAIND (62 cases), DBK (709 cases), DBKND (2 cases), DEL (1917 cases), DELND (5 cases), FAV (8683 cases).

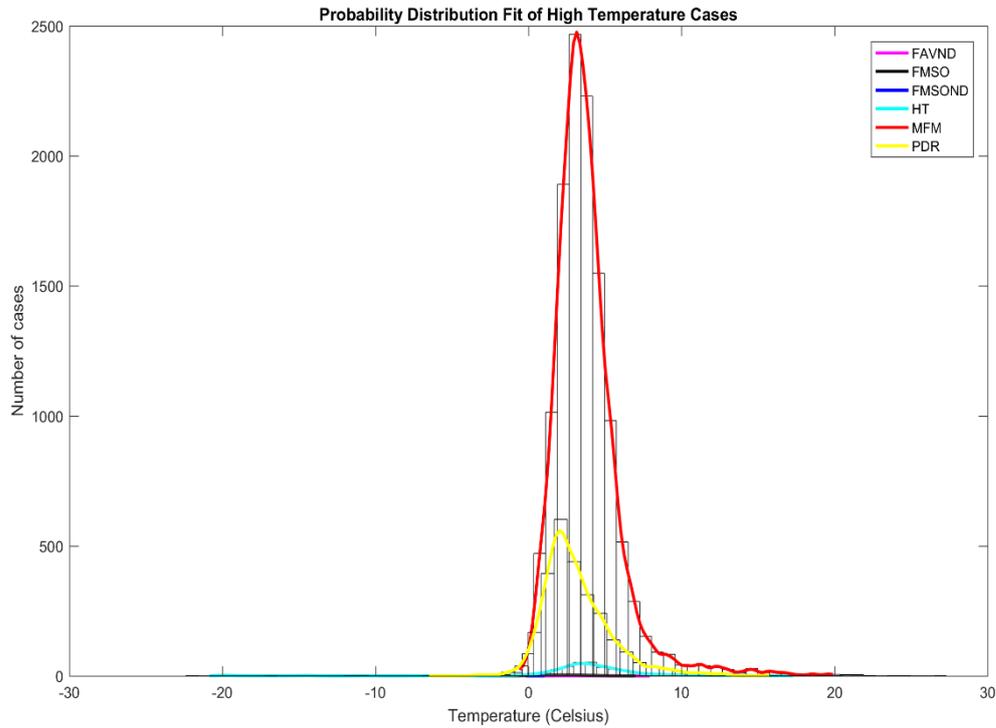

Figure 15. Number of cases and CPT probability distributions of the following types of HT cases: FAVND (19 cases), FMSO (109 cases), FMSOND (2 cases), HT (335 cases), MFM (1213 cases), PDR (2622).

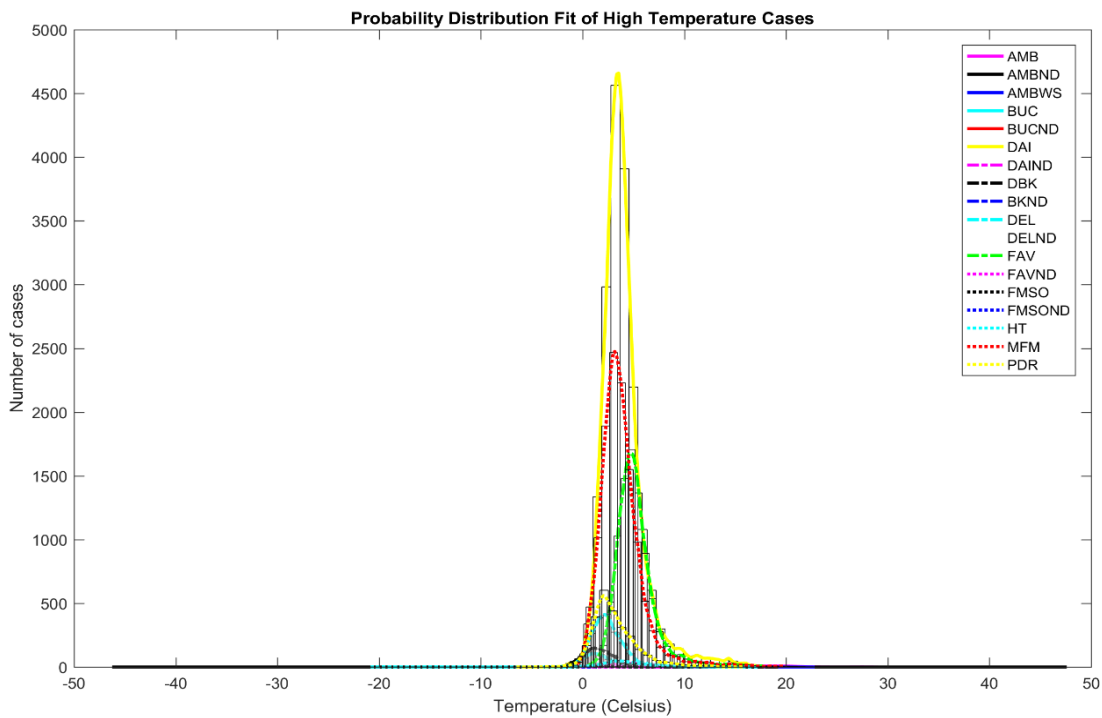

Figure 16. Number of cases and CPT probability distributions of different types of HT cases.

The highest number of HT cases is in order DAI (18226 cases), MFM (12131 cases), PDR (2622 cases), FAV (8683 cases) and DEL (1917 cases).

## B. *Longer period of time of several minutes*

Data (temperature, status) from the estate of 63930 cases from a longer period of time of 10 minutes is processed. The 10 consecutive minutes are the following : 0, 1, 2, 3, 4, 5, 6, 7, 8, 9, 10. Also, 584 cases have been excluded as they have missing data which results in 63346 cases which are available.

Moreover, not all the cases have data for the duration of the 10 minutes (i.e. starting at minute 0 and ending to minute 10). Therefore, a polynomial interpolation method is used in order to cover the full duration of 10 minutes. For example, a HT refrigeration case from a given store has data for the minutes 1, 4, 6, 8 and 10 minute with the values 3, 3, 2.9, 2.9 and 2.9 degrees Celsius. A polynomial interpolation method of second degrees is used and the resulted values (i.e. 3.0238, 3.0059, 2.9891, 2.9732, 2.9583, 2.9444, 2.9315, 2.9196, 2.9086, 2.8987, 2.8897) are shown in Figure 17a. Another LT refrigeration case from a different store has data for the minutes 0, 3, 6 and 9 minute with the values -20.7, -20.7, -20.7 and -20.7 degrees Celsius. Again, a polynomial interpolation method of second degrees is used and the resulted values representing a straight line (i.e. -20.7, -20.7, -20.7, -20.7, -20.7, -20.7, -20.7, -20.7, -20.7, -20.7, -20.7) are shown in Figure 17b. Figure 18 shows the temperature evolution for 10 minutes for 63346 cases, both HT and LT cases.

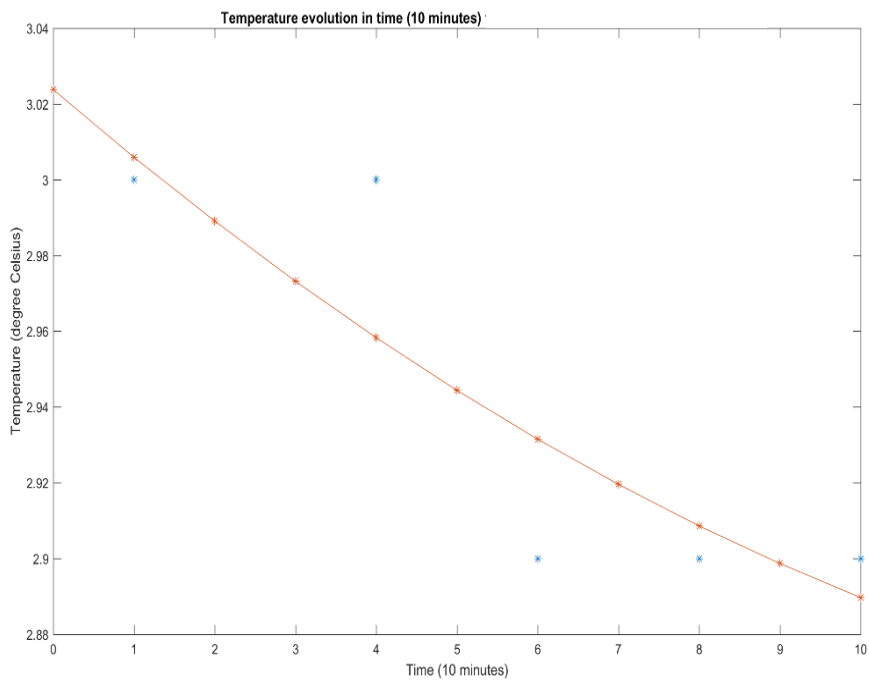

a) CPT temperature evolution in time (10 minutes) for a HT refrigeration case from a store (blue stars – original data; red line – interpolated data).

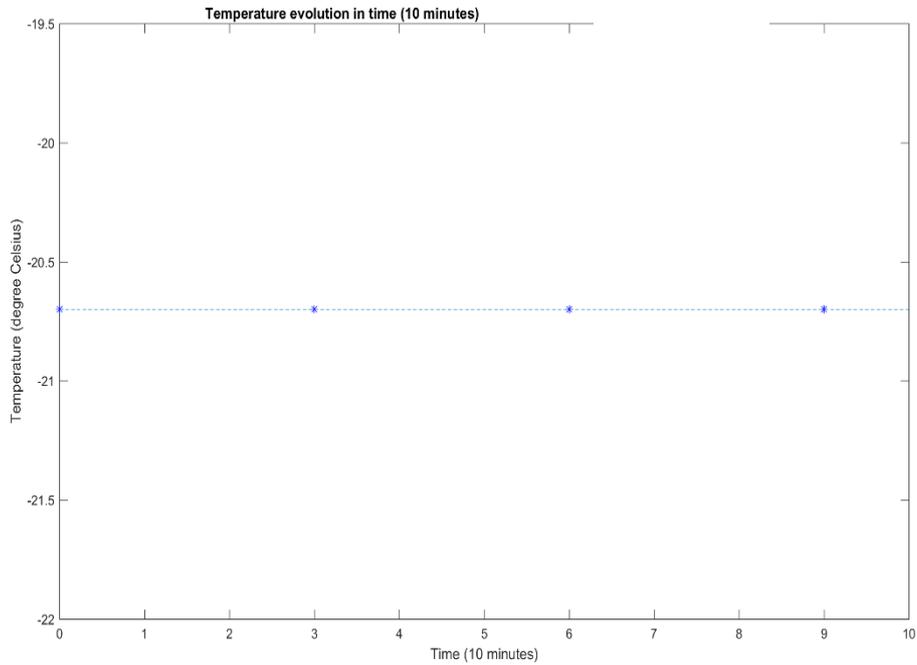

b) CPT temperature evolution in time (10 minutes) for a LT case (i.e. refrigeration case) from a second store (blue star – original data, dashed line – interpolated data).

Figure 17. CPT temperature evolution in time (10 minutes) for two refrigeration cases.

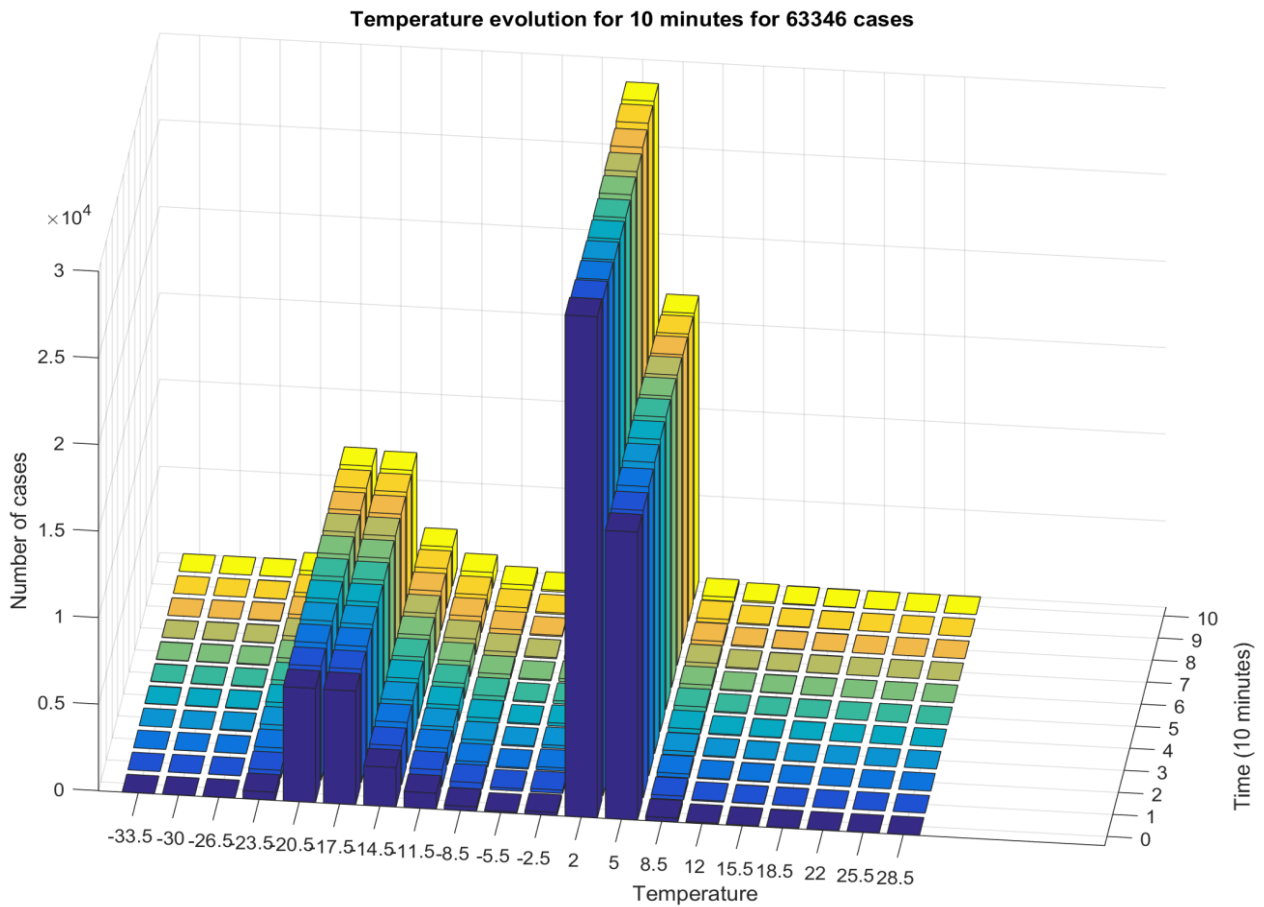

Figure 18. CPT temperature evolution for 10 minutes for 63346 cases including HT and LT cases and real data.

## C. Determination of warm-up rates for a single LT case and a single HT case from a commercial retail shop

Yearly data from cases from a store was used to study the evolution of CPT over a year. The cases were both LT and HT cases.

The warm-up rate is usually calculated as following:

$$Warm-up\ rate = \frac{Temp_f - Temp_i}{Time_f - Time_i} \quad (1)$$

where $Temp_f$ is the CPT at the end of the defrost, $Temp_i$ is the initial CPT at the beginning of the defrost ($Temp_f \geq Temp_i$), and $Time_f$ and $Time_i$ are the starting and ending time of the defrost and measured in minutes. Similarly, a cold-down rate can be defined.

The CPT for a LT case for 144 days is shown in Figure 19. In Figure 20 is shown again for the same refrigeration case, the Air-off temperature and the CPT for the same LT case for 250 minutes. The spike before the 50 minutes time means that a defrost takes place in Figure 20. Typically, a warm-up rate of 0.2 deg/min is used for the CPT during the defrost for this LT case.

It is also of interest the calculation of a cold down rate for the CPT, which can be derived from Figure 20 and Figure 21. At time 45 minute the CPT is about -16 degrees Celsius while at time 230 minute the CPT is about -18 degrees Celsius, which means a cold down rate for CPT of 0.0108 degrees Celsius per minute. The cold down rate in this case 0.0108 degrees Celsius per minute is much slower than the warm up rate, which was 0.2 degrees Celsius per minute.

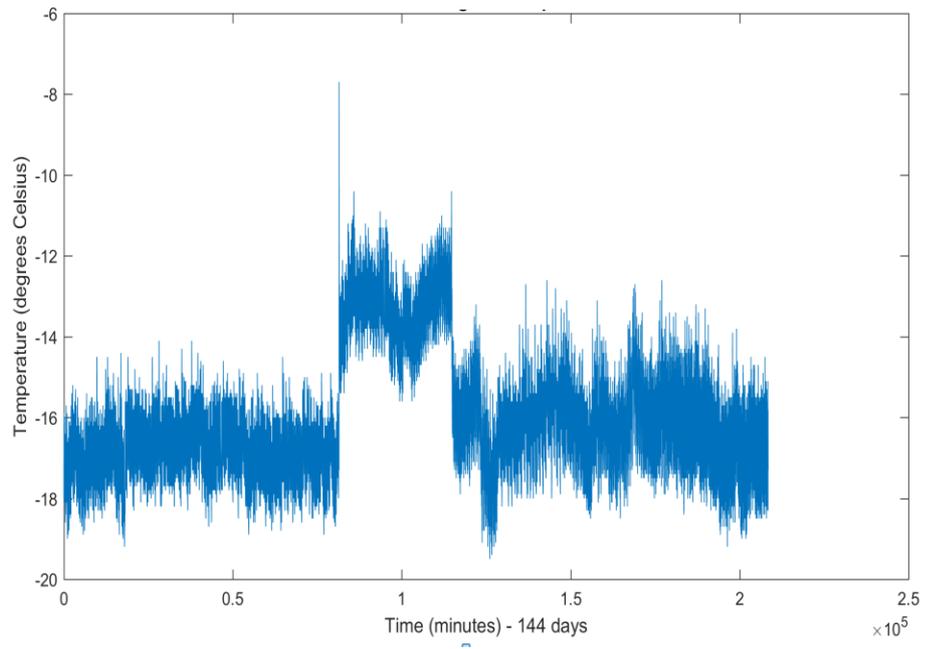

Figure 19. CPT for a LT case for 144 days.

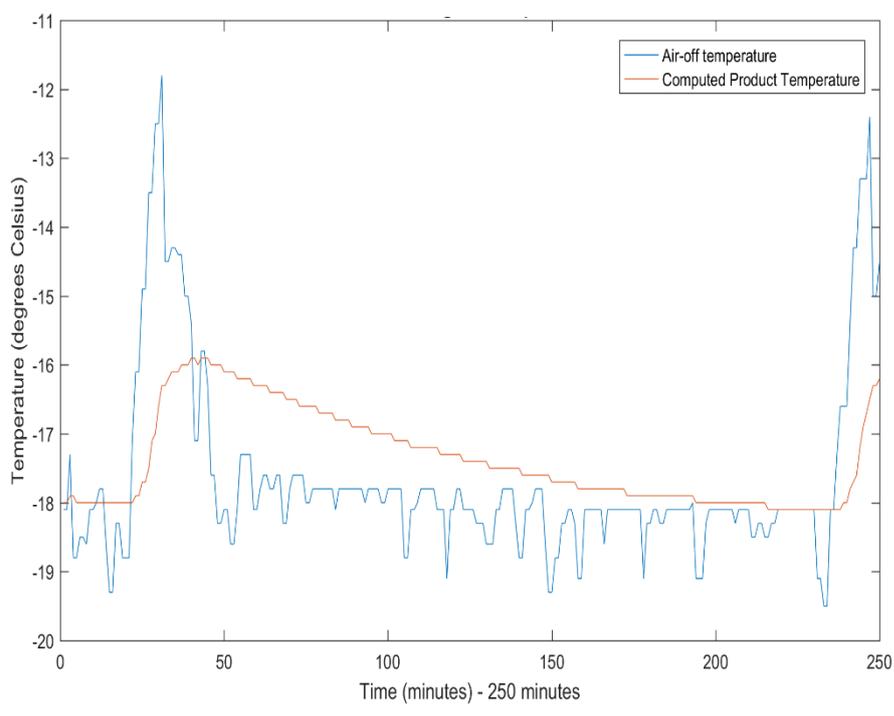

Figure 20. Air-off temperature and CPT for a LT case for 250 minutes.

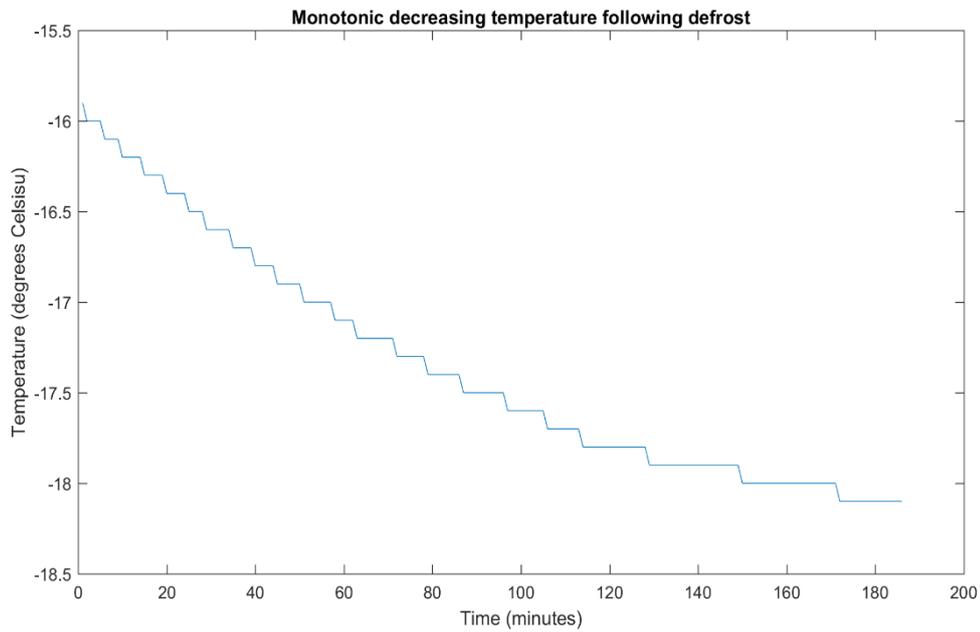

Figure 21. CPT for a LT case following defrost and with a monotonic decreasing slope.

The CPT for a HT case for 249 days is shown in Figure 22. Figure 23 shows again for the same case, the Air-on temperature, the Air-off temperature and the CPT for the same HT case for 24 hours. The spikes shown in Figure 23 correspond to defrosts. Typically, a warm-up rate of 0.05 degrees Celsius per minute is used for the CPT during the defrost.

It is also of interest to calculate a cold down rate for the CPT, which can be derived from Figure 24. At time 160 minute the CPT is about 6.7 degrees Celsius while at time 320 minute the CPT is about 4.4 degrees Celsius, which means a cold down rate for CPT of 0.0144 degrees Celsius per minute. The cold down rate in this case 0.0144 degrees Celsius per minute is much slower than the warm up rate, which was 0.05 degrees Celsius per minute.

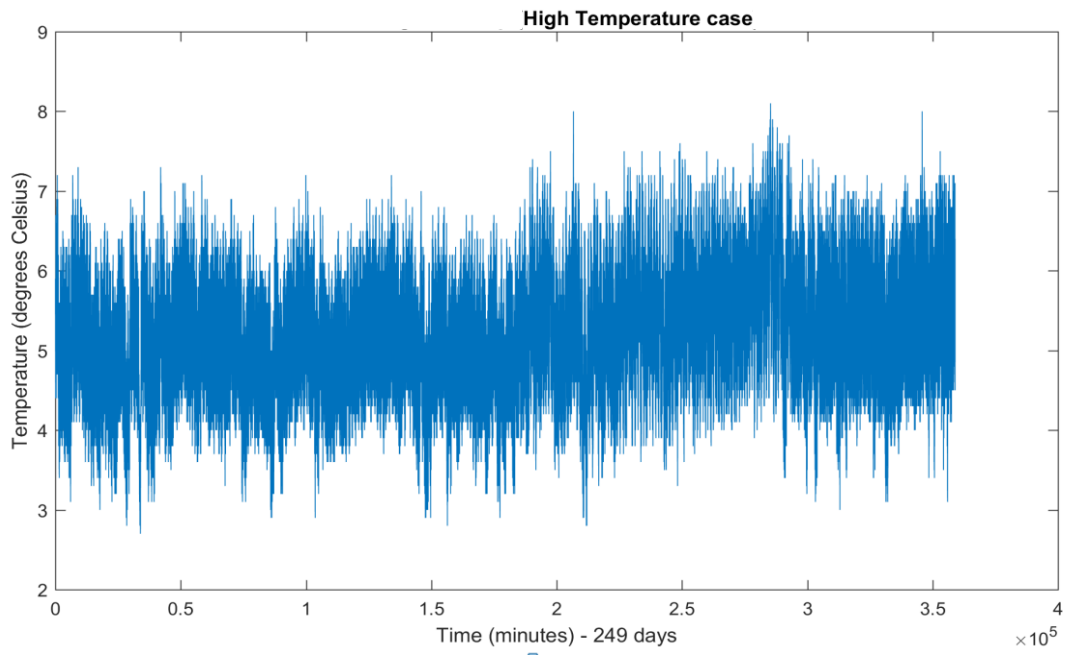

Figure 22. CPT for a HT case for 249 days.

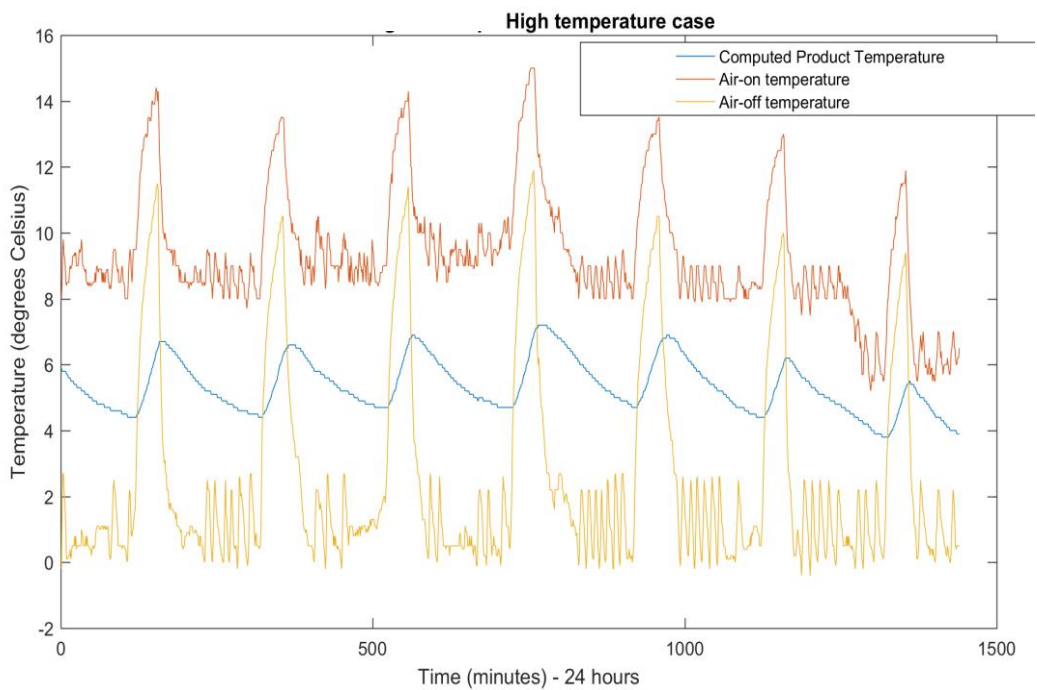

Figure 23. Air-on temperature, Air-off temperature and CPT for a HT case for 24 hours.

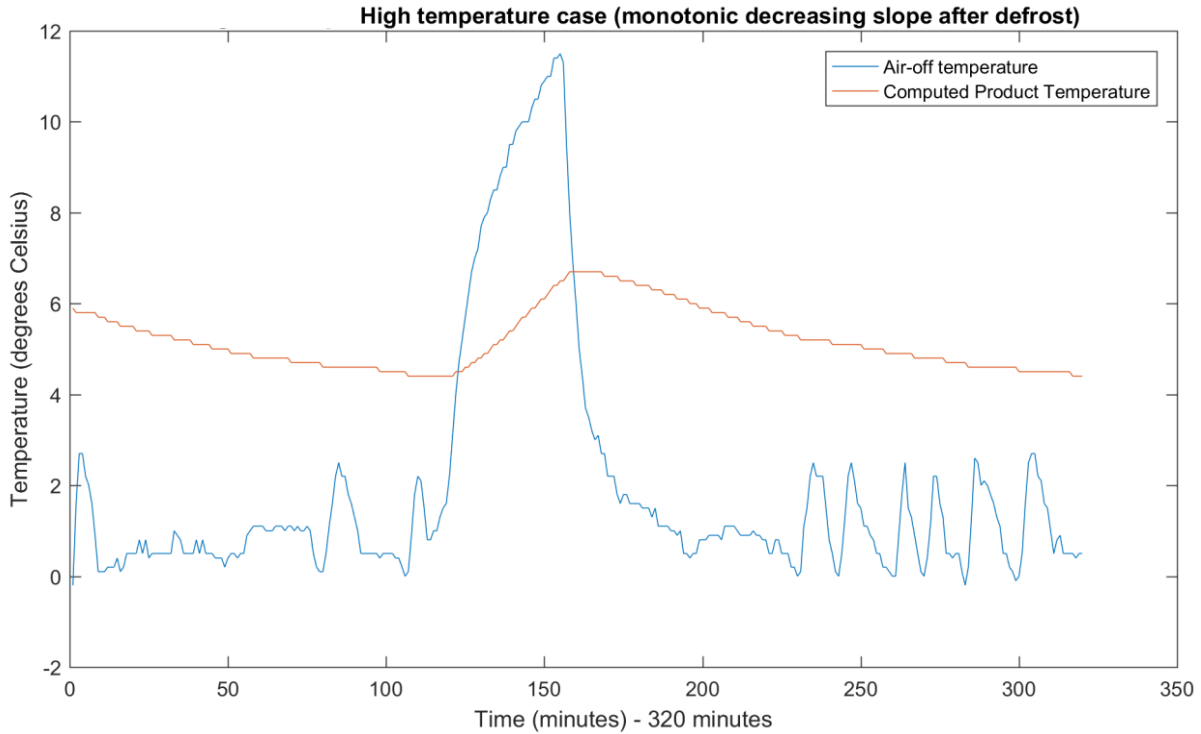

Figure 24. Air-off temperature and CPT for a HT case for 320 minutes with monotonic decreasing slope after defrost.

*D. Determination of warm-up temperature rates for multiple HT and LT cases in multiple shops and throughout a year*

Data from multiple shops is available including the CPT and the Defrost status between at least 290 days and a maximum of 1 year.

A first shop is again investigated and a LT case (Figure 25) is studied. In total there are 289 days available and a total of 572 defrosts were counted and 4 defrosts were excluded because of impracticalities in reading properly the respective data at the present time. The mean for a normal probability distribution associated to this warm-up rate would be 0.1081 degrees Celsius per minute and a standard distribution of 0.0118 degrees Celsius per minute, which corresponds to the values from the literature. A maximum rate of 0.1548 degrees Celsius per minute was noticed.

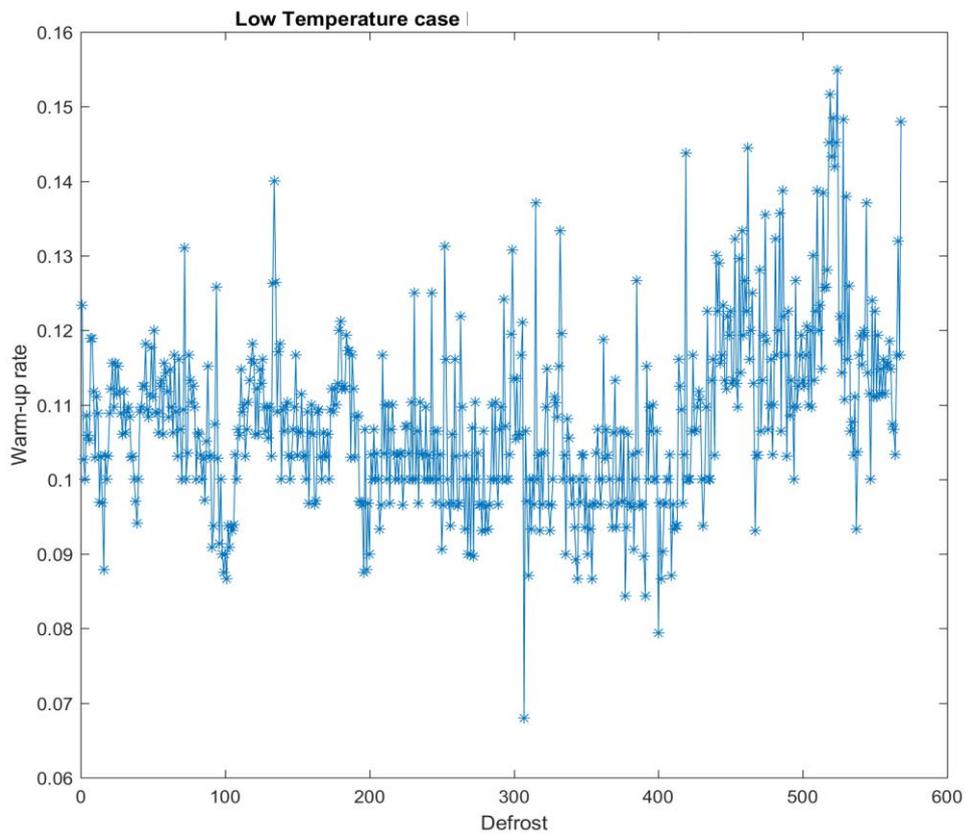

Figure. 25. Warm-up rates (degrees Celsius per minute) for a LT case.

Second, a HT case is studied (Figure 26). In total there were 289 days available and 3117 defrosts were counted. The mean for a normal probability distribution associated to this warm-up rate would be 0.0291 degrees Celsius per minute and a standard distribution of 0.0063 degrees Celsius per minute. A maximum warm-up rate of 0.0556 degrees Celsius per minute was noticed.

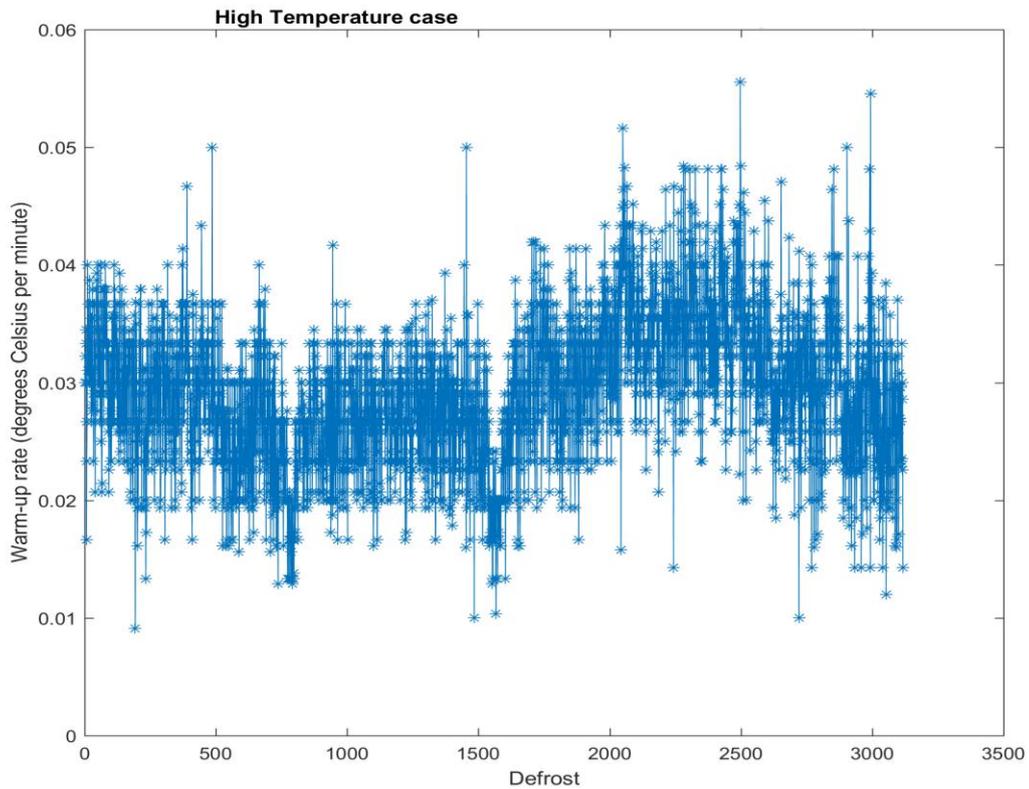

Figure. 26. Warm-up rates (degrees Celsius per minute) for a HT case.

For this first shop, Table 2 and Table 3 summarize for 70 cases comprising 52 HT cases and 18 LT cases, the means and the standard distributions of the warm-up rates of the respective cases over a period of time approximately ranging between 289 days and 1 year and consisting in total of 10000s of defrosts cycles.   Overall there is a mean of 0.1520 degrees Celsius per minute and a standard deviation of 0.0198 degrees Celsius per minute for the LT cases.    For the HT cases there is a mean of 0.0339 degrees Celsius per minute and a standard deviation of 0.0071 degrees Celsius per minute. Few cases were excluded because either the cases did not seem to have a defrost status or data was too messy or defrost data too short (i.e. less than a minute).

| Case | Mean | Standard deviation |
|------|------|--------------------|
| 1    | 0.1163 | 0.0166 |
| 2    | 0.1027 | 0.0340 |
| 3    | 0.1771 | 0.0199 |
| 45   | 0.1450 | 0.0187 |
| 6    | 0.1604 | 0.0121 |
| 7    | 0.1591 | 0.0143 |
| 8    | 0.1812 | 0.0215 |
| 9    | 0.1910 | 0.0135 |
| 10   | 0.1392 | 0.0202 |
| 11   | 0.2056 | 0.0170 |
| 12   | 0.1512 | 0.0222 |
| 13   | 0.1559 | 0.0199 |

| Case | Mean | Standard deviation |
|---|---|---|
| 14 | 0.1310 | 0.0233 |
| 15 | 0.1319 | 0.0196 |
| 16 | 0.1436 | 0.0222 |
| 17 | 0.1460 | 0.0173 |
| 18 | 0.1387 | 0.0292 |
| 19 | 0.1598 | 0.0148 |

Table 2. Warm up rates - mean and standard deviations of LT cases (degrees Celsius per minute).

| Case | Mean | Standard deviation |
|---|---|---|
| 1 | 0.0411 | 0.0234 |
| 2 | 0.0340 | 0.0187 |
| 3 | 0.0346 | 0.0165 |
| 4 | 0.0317 | 0.0154 |
| 5 | 0.0340 | 0.0166 |
| 6 | 0.0324 | 0.0102 |
| 7 | 0.0281 | 0.0048 |
| 8 | 0.0321 | 0.0052 |
| 9 | 0.0394 | 0.0050 |
| 10 | 0.0416 | 0.0058 |
| 11 | 0.0297 | 0.0056 |
| 12 | 0.0559 | 0.0091 |
| 13 | 0.0282 | 0.0168 |
| 14 | 0.0317 | 0.0063 |
| 15 | 0.0326 | 0.0066 |
| 16 | 0.0301 | 0.0064 |
| 17 | 0.0389 | 0.0064 |
| 18 | 0.0332 | 0.0064 |
| 19 | 0.0311 | 0.0054 |
| 20 | 0.0382 | 0.0067 |
| 21 | 0.0298 | 0.0056 |
| 22 | 0.0372 | 0.0054 |
| 23 | 0.0321 | 0.0054 |
| 24 | 0.0368 | 0.0072 |
| 25 | 0.0331 | 0.0067 |
| 26 | 0.0294 | 0.0056 |
| 27 | 0.0312 | 0.0071 |
| 28 | 0.0284 | 0.0065 |
| 29 | 0.0266 | 0.0063 |
| 30 | 0.0323 | 0.0054 |
| 31 | 0.0313 | 0.0053 |
| 32 | 0.0221 | 0.0063 |
| 33 | 0.0285 | 0.0052 |
| 34 | 0.0369 | 0.0050 |
| 35 | 0.0302 | 0.0058 |
| 36 | 0.0255 | 0.0063 |
| 37 | 0.0450 | 0.0309 |
| 38 | 0.0337 | 0.0049 |
| 39 | 0.0325 | 0.0050 |
| 40 | 0.0451 | 0.0061 |
| 41 | 0.0373 | 0.0057 |
| 42 | 0.0447 | 0.0058 |
| 43 | 0.0395 | 0.0058 |
| 44 | 0.0405 | 0.0058 |
| 45 | 0.0415 | 0.0053 |
| 46 | 0.0317 | 0.0051 |
| 47 | 0.0689 | 0.0256 |
| 48 | 0.0309 | 0.0071 |
| 49 | 0.0294 | 0.0057 |
| 50 | 0.0307 | 0.0065 |
| 51 | 0.0312 | 0.0062 |
| 52 | | |

Table 3. Warm up rates - mean and standard deviations of HT cases (degrees Celsius per minute).

A second shop, which has 115 cases, is processed. The data covers 1 year. There are 31 LT cases (Table 4) and 84 HT cases (Table 5). There is a mean for the warm-up rates for LT cases of 0.1526 and a standard deviation of 0.0156 degrees Celsius per minute. There is also a mean for the warm-up rates for HT cases of 0.0364 and a standard deviation of 0.0080 degrees Celsius per minute. There is some important variation through all the LT and HT cases. The HT cases with a double star did not have defrost status available in the dataset.

| Case | Mean | Standard deviation |
|---|---|---|
| 1 | 0.1336 | 0.0087 |
| 2 | 0.1152 | 0.0072 |
| 3 | 0.1348 | 0.0165 |
| 4 | 0.1771 | 0.0274 |
| 5 | 0.1248 | 0.0175 |
| 6 | 0.1525 | 0.0199 |
| 7 | 0.2034 | 0.0113 |
| 8 | 0.1396 | 0.0099 |
| 9 | 0.1387 | 0.0174 |
| 10 | 0.1234 | 0.0140 |
| 11 | 0.1416 | 0.0103 |
| 12 | 0.1500 | 0.0228 |
| 13 | 0.1237 | 0.0161 |
| 14 | 0.1175 | 0.0099 |
| 15 | 0.0961 | 0.0131 |
| 16 | 0.2185 | 0.0144 |
| 17 | 0.2174 | 0.0152 |
| 18 | 0.1354 | 0.0141 |
| 19 | 0.1418 | 0.0127 |
| 20 | 0.1362 | 0.0137 |
| 21 | 0.1817 | 0.0212 |
| 22 | 0.1626 | 0.0143 |

| Case | Mean | Standard deviation |
|---|---|---|
| 23 | 0.1163 | 0.0191 |
| 24 | 0.1897 | 0.0160 |
| 25 | 0.1425 | 0.0191 |
| 26 | 0.1188 | 0.0108 |
| 27 | 0.1751 | 0.0156 |
| 28 | 0.1284 | 0.0117 |
| 29 | 0.2037 | 0.0217 |
| 30 | 0.2078 | 0.0158 |
| 31 | 0.1819 | 0.0253 |

Table 4. Warm-up rates - mean and standard deviations of LT cases (degrees Celsius per minute).

| Case | Mean | Standard deviation |
|---|---|---|
| 1 | 0.0309 | 0.0052 |
| 2 | 0.0306 | 0.0048 |
| 3 | 0.0302 | 0.0043 |
| 4 | 0.0243 | 0.0042 |
| 5 | 0.0339 | 0.0054 |
| 6 | 0.0353 | 0.0063 |
| 7 | 0.0323 | 0.0066 |
| 8 | 0.0290 | 0.0056 |
| 9 | 0.0402 | 0.0107 |
| 10 | 0.0313 | 0.0059 |
| 11 | 0.0406 | 0.0079 |
| 12 | 0.0368 | 0.0114 |
| 13 | 0.0305 | 0.0050 |
| 14 | 0.0413 | 0.0069 |
| 15 | 0.0312 | 0.0061 |
| 16 | 0.0331 | 0.0067 |
| 17 | 0.0445 | 0.0084 |
| 18 | 0.0412 | 0.0108 |
| 19 | 0.0392 | 0.0071 |
| 20 | 0.0348 | 0.0064 |
| 21 | 0.0348 | 0.0068 |
| 22 | 0.0317 | 0.0036 |
| 23 | 0.0453 | 0.0053 |
| 24 | 0.0259 | 0.0058 |
| 25 | 0.0285 | 0.0056 |
| 26 | 0.0390 | 0.0079 |
| 27 | 0.0305 | 0.0058 |
| 28 | 0.0305 | 0.0054 |
| 29 | 0.0325 | 0.0048 |
| 30 | 0.0293 | 0.0059 |
| 31 | 0.0369 | 0.0048 |
| 32 | 0.0405 | 0.0066 |
| 33 | 0.0493 | 0.0072 |
| 34 | 0.0483 | 0.0081 |
| 35 | 0.0442 | 0.0078 |
| 36 | 0.0482 | 0.0069 |
| 37 | 0.0353 | 0.0072 |
| 38 | 0.0655 | 0.0122 |
| 39** | | |
| 40 | 0.0278 | 0.0050 |
| 41 | 0.0900 | 0.0496 |
| 42 | 0.0879 | 0.0178 |
| 43 | 0.0988 | 0.0273 |
| 44 | 0.0842 | 0.0211 |
| 45** | | |
| 46 | 0.0399 | 0.0246 |
| 47 | 0.0528 | 0.0331 |
| 48 | 0.0590 | 0.0189 |
| 49 | 0.0424 | 0.0057 |
| 50 | 0.0327 | 0.0064 |
| 51 | 0.0309 | 0.0062 |
| 52 | 0.0197 | 0.0053 |
| 53 | 0.0321 | 0.0045 |
| 54 | 0.0230 | 0.0061 |
| 55 | 0.0312 | 0.0077 |
| 56 | 0.0470 | 0.0063 |
| 57 | 0.0362 | 0.0062 |
| 58 | 0.0279 | 0.0063 |
| 59 | 0.0143 | 0.0043 |
| 60 | 0.0071 | 0.0051 |
| 61 | 0.0205 | 0.0060 |
| 62 | 0.0367 | 0.0063 |
| 63 | 0.0252 | 0.0093 |
| 64 | 0.0266 | 0.0055 |
| 65 | 0.0368 | 0.0137 |
| 66 | 0.0279 | 0.0056 |
| 67 | 0.0244 | 0.0065 |
| 68 | 0.0324 | 0.0053 |
| 69 | 0.0533 | 0.0082 |
| 70 | 0.0235 | 0.0043 |
| 71 | 0.0240 | 0.0038 |
| 72 | 0.0259 | 0.0037 |
| 73 | 0.0290 | 0.0041 |
| 74 | 0.0223 | 0.0051 |
| 75 | 0.0200 | 0.0033 |
| 76 | 0.0299 | 0.0038 |
| 77 | 0.0201 | 0.0042 |
| 78 | 0.0181 | 0.0036 |
| 79 | 0.0419 | 0.0059 |
| 80 | 0.0450 | 0.0076 |
| 81 | 0.0302 | 0.0053 |
| 82 | 0.0286 | 0.0062 |
| 83 | 0.0309 | 0.0053 |
| 84 | 0.0392 | 0.0061 |

Table 5. Warm-up rates - mean and standard deviations of HT cases (degrees Celsius per minute).

There is a third shop being investigated which has 50 cases to be processed. The data covers 1 year. There are 12 LT cases (Table 6) and 38 HT cases (Table 7). There is a mean for the warm-up rates for LT cases of 0.1523 degrees Celsius per minute and a standard deviation of 0.0202 degrees Celsius per minute. There is also a mean for the warm-up rates for HT cases of 0.0401 and a standard deviation of 0.0063 degrees Celsius per minute.

| Case | Mean | Standard deviation |
| --- | --- | --- |
| 1 | 0.0289 | 0.0045 |
| 2 | 0.0904 | 0.0168 |
| 3 | 0.0188 | 0.0133 |
| 4 | 0.0967 | 0.0115 |
| 5 | 0.0426 | 0.0051 |
| 6 | 0.0286 | 0.0044 |
| 7 | 0.0416 | 0.0047 |
| 8 | 0.0340 | 0.0040 |
| 9 | 0.0384 | 0.0042 |
| 10 | 0.0372 | 0.0046 |
| 11 | 0.0355 | 0.0042 |
| 12 | 0.0395 | 0.0051 |
| 13 | 0.0403 | 0.0054 |
| 14 | 0.0432 | 0.0047 |
| 15 | 0.0400 | 0.0038 |
| 16 | 0.0360 | 0.0071 |
| 17 | 0.0373 | 0.0041 |
| 18 | 0.0430 | 0.0052 |
| 19 | 0.0363 | 0.0034 |
| 20 | 0.0333 | 0.0033 |
| 21 | 0.0354 | 0.0031 |
| 22 | 0.0525 | 0.0060 |
| 23 | 0.0839 | 0.0037 |
| 24 | 0.0274 | 0.0194 |
| 25 | 0.0360 | 0.0147 |
| 26 | 0.0374 | 0.0044 |
| 27 | 0.0283 | 0.0049 |
| 28 | 0.0294 | 0.0057 |
| 29 | 0.0526 | 0.0104 |
| 30 | 0.0212 | 0.0063 |
| 31 | 0.0376 | 0.0055 |
| 32 | 0.0384 | 0.0066 |
| 33 | 0.0358 | 0.0046 |
| 34 | 0.0297 | 0.0050 |
| 35 | 0.0315 | 0.0044 |
| 36 | 0.0273 | 0.0048 |
| 37 | 0.0368 | 0.0042 |
| 38 | 0.0392 | 0.0051 |

Table 6. Warm-up rates - mean and standard deviations of HT cases (degrees Celsius per minute).

| Case | Mean | Standard deviation |
|------|--------|--------------------|
| 1    | 0.0600 | 0.0223 |
| 2    | 0.0762 | 0.0158 |
| 3    | 0.0559 | 0.0147 |
| 4    | 0.1929 | 0.0185 |
| 5    | 0.1842 | 0.0185 |
| 6    | 0.1664 | 0.0232 |
| 7    | 0.1824 | 0.0192 |
| 8    | 0.1897 | 0.0202 |
| 9    | 0.1850 | 0.0221 |
| 10   | 0.1839 | 0.0261 |
| 11   | 0.1743 | 0.0205 |
| 12   | 0.1763 | 0.0218 |

Table 7. Warm-up rates - mean and standard deviations of LT cases (degrees Celsius per minute).

To further understand these warm-up rates and cold-down rates, furthermore Figure 27 shows the CPT evolution for 68721 cases/fridges during 24 hours of a day during March month for HT and LT cases (i.e. hourly data).

Especially for the HT cases in Figure 27b, it can be seen clearly that during night between 23 o'clock and 7 o'clock when the CPT of the HT cases decreases then the lengths of the long green bars (i.e. the total number of HT cases with the CPT around 2 degrees Celsius) are increasing in size. By contrast in Figure 27b the CPT during the afternoon increases for the HT cases, as it is possible to see that the lengths of the long blue bars are decreasing, and the respective long blue bars correspond to the total number of fridges/cases with the CPT around 2 degrees Celsius.

Figure 27b shows also some slim lines, which are representing the total number of HT cases with the CPT around 5 degrees Celsius: it can be observed that these HT cases with the CPTs around 5 degrees Celsius are increasing their numbers during the afternoon and are decreasing their numbers during the night when the CPT of the HT cases/fridges is decreasing to around 2 degrees Celsius.

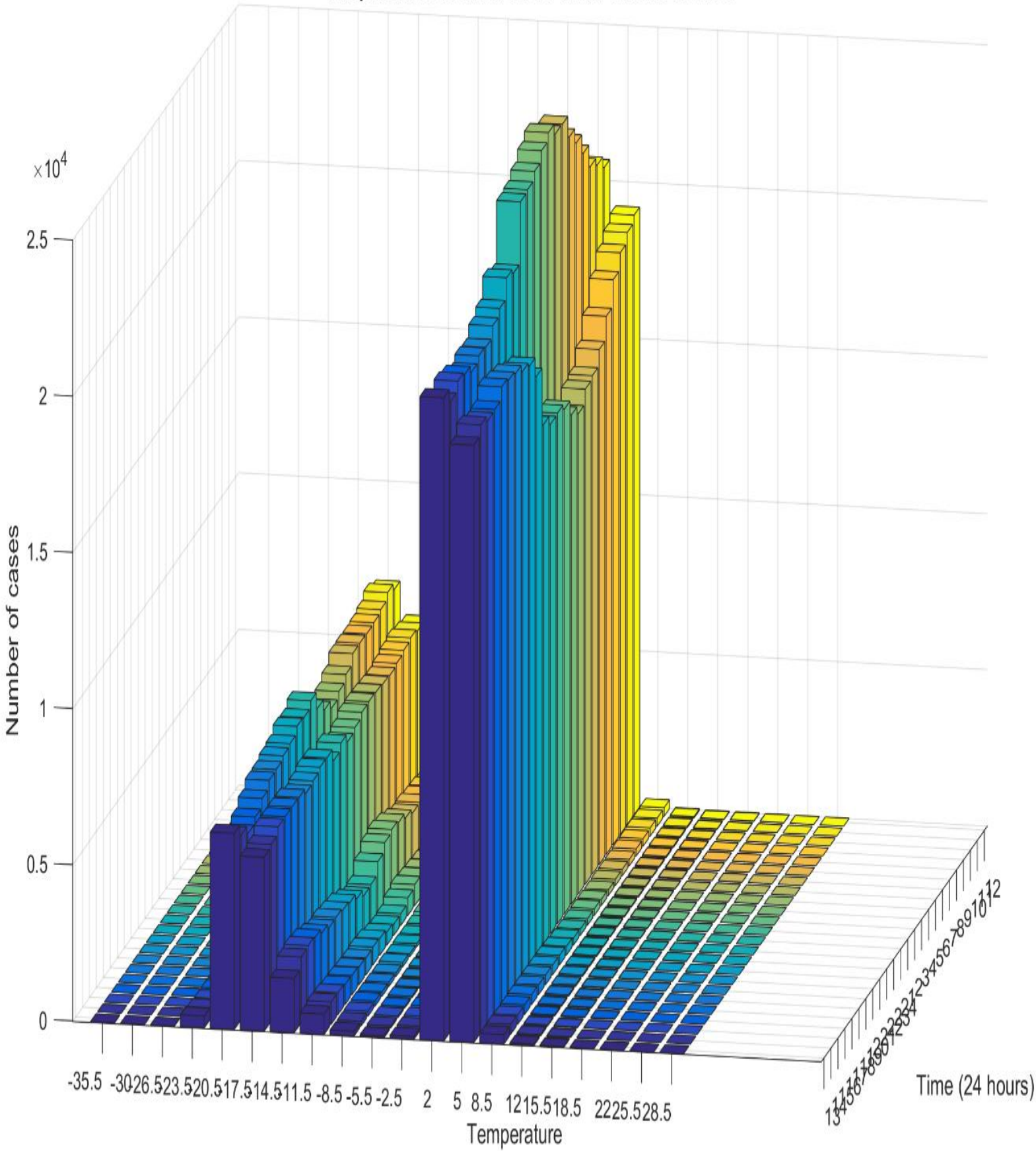

a) CPT (degrees Celsius) for HT and LT cases during the 24 hours period of time.

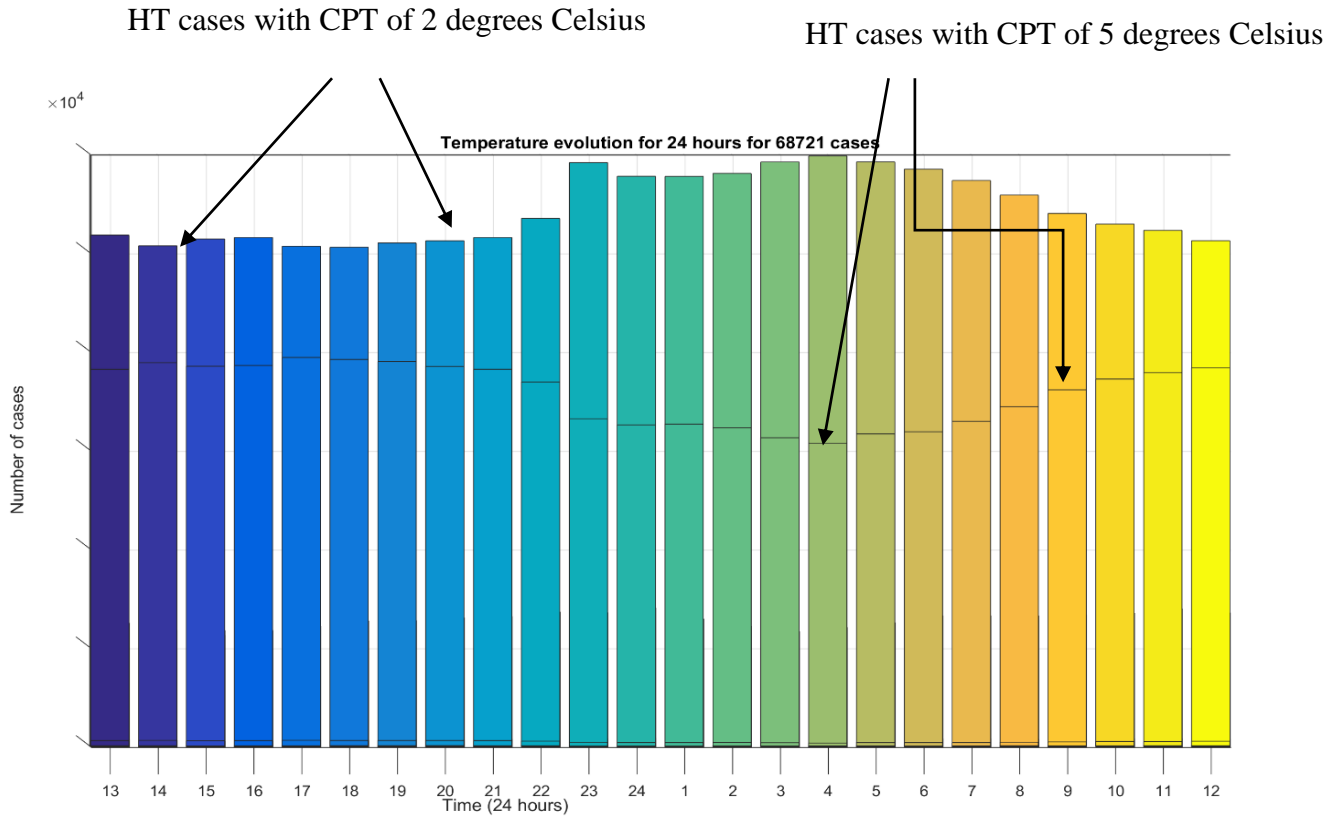

b) Number of HT fridges/cases with CPT around 2 degrees Celsius and 5 degrees Celsius for 68721 fridges/cases during 24 hours:

- during night between 23 o'clock and 7 o'clock the CPT of the HT cases decreases as the long green bars are the HT cases with the CPT around 2 degrees Celsius; by contrast, the CPT during the afternoon (between 12 o'clock and 18 o'clock) increases for the HT cases and this can be noticed in the lengths of the long blue bars which are decreasing as the respective long blue bars correspond to the fridges/cases with the CPT around 2 degrees Celsius;

- in the same second figure (b), it can be observed some slim lines, which are representing the number of HT cases with the CPT around 5 degrees Celsius; it can be observed the opposite effect, that these HT cases with the CPT around 5 degrees Celsius are increasing their numbers during the afternoon (between 12 o'clock and 18 o'clock) and are decreasing their numbers during the night (between 23 o'clock and 7 o'clock).

Figure 27. HT and LT cases during 24 hours operational period of time.

The following cases/fridges were selected as being the HT cases/fridges totalizing 49897 cases: 'MFM', 'HT', 'AMB', 'DAI', 'PDR', 'FAV', 'BWS', 'BUC', 'DEL', 'DAIND', 'DBK', 'FMSO', 'AMBND', 'FAVND', 'BUCND', 'DELND', 'FMSOND', 'DBKND'.

Figure 27b is shown again in Figure 28 at a higher resolution for HT cases.

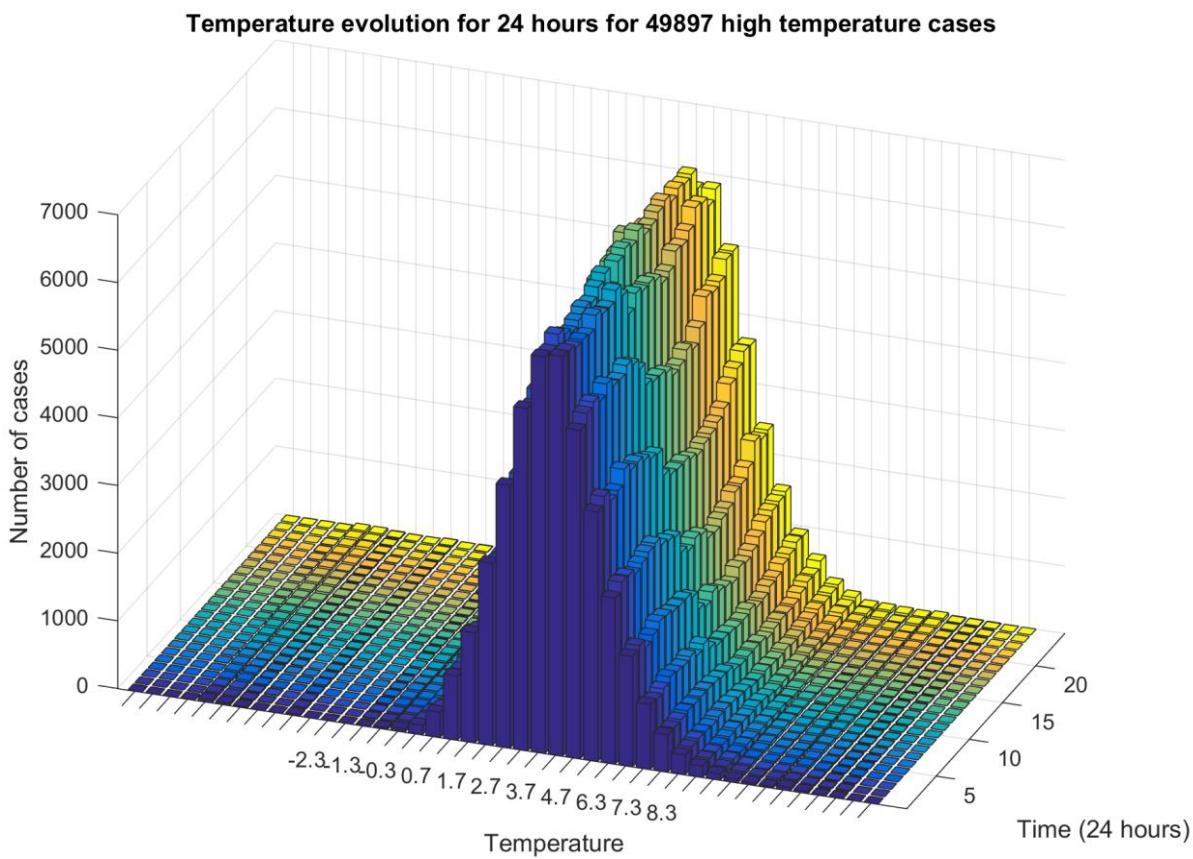

Figure. 28. Number of fridges/cases (41730) with CPT temperature around 2 degrees Celsius and 5 degrees Celsius for 68721 fridges/cases during 24 hours (i.e. higher resolution than Figure 27b) during 24 hours period of time.

The cold fridges/cases are called as FFW, FFF, FFGD, LT, FFGDND, FFWND, FFFND. There are in total 18653 LT fridges/cases. Figure 29 shows the total number of LT fridges/cases (18653) with the CPT around -20 degrees Celsius and -17 degrees Celsius for 18653 fridges/cases during 24 hours and the same figure is shown again in Figure 30 from a different perspective. It can be observed clearly that the CPT of the LT cases decreases during the night and early afternoon, that is between hours 1 o'clock to 5 o'clock and also during early afternoon between 13 o'clock and 15 o'clock when people are at work. By opposite, during evening, that is between 17 o'clock and 21 o'clock (i.e. people finished work and some of them are doing shopping) and also during the morning, that is between 9 o'clock and 12 o'clock, the CPT increases as it can be seen in Figure 30.

The second spike could be very likely because of the defrost which tends to converge around a value of -17 degrees Celsius.

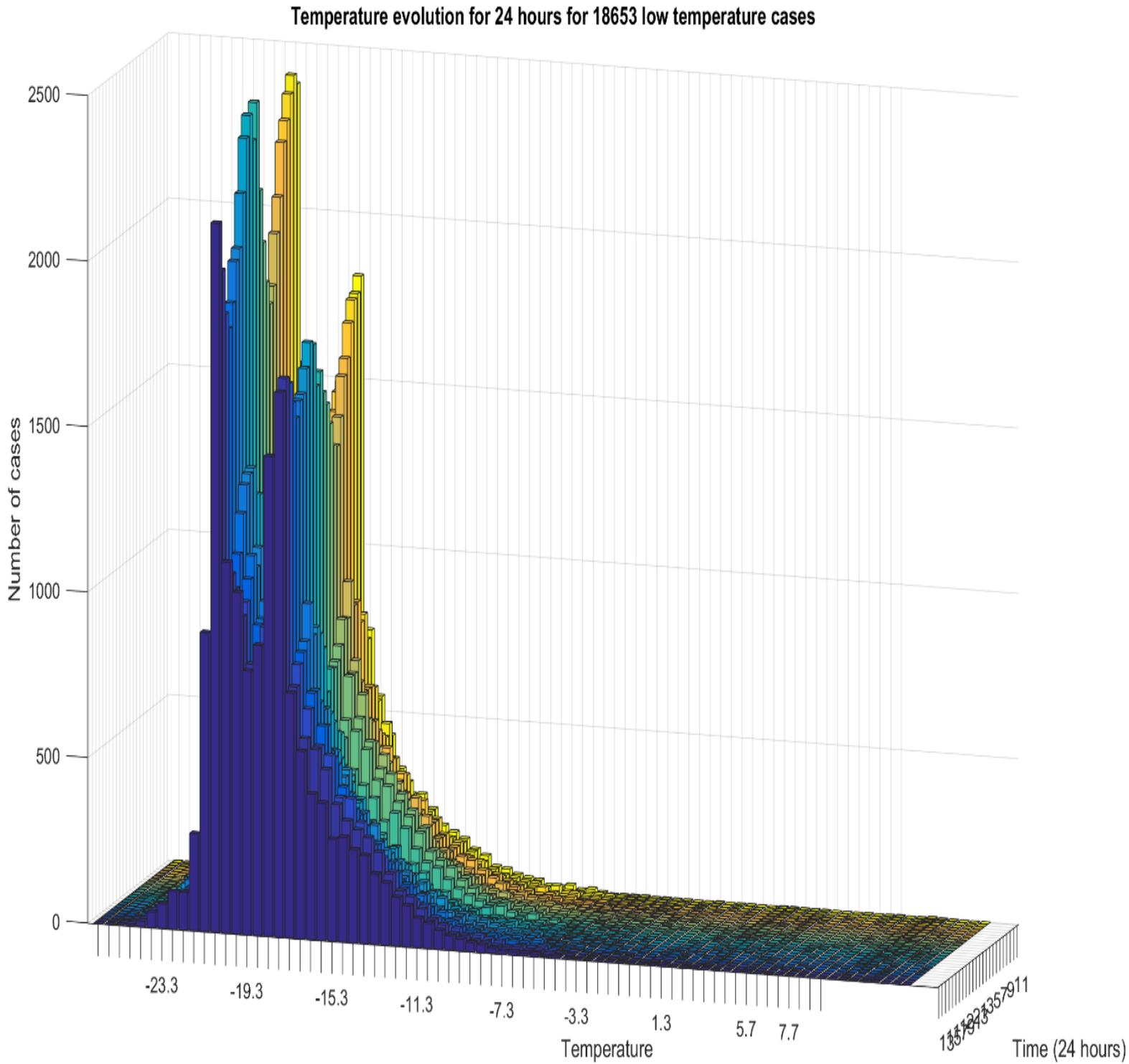

Figure 29. Number of LT fridges/cases (18653) with CPT around -20 degrees Celsius and -17 degrees Celsius for 18653 fridges/cases during 24 hours.

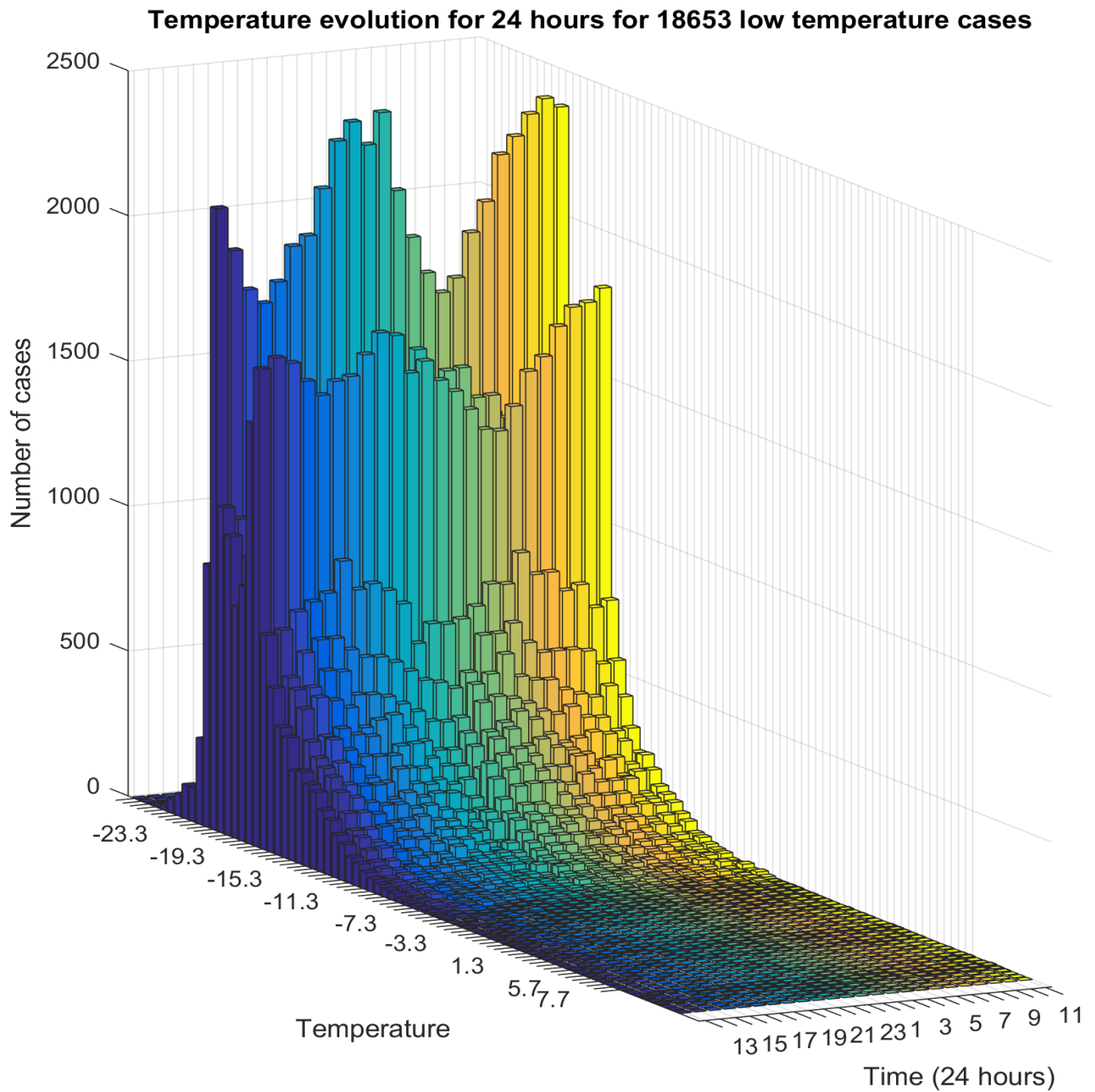

Figure. 30. Figure 29 shown from a different perspective : total number of LT fridges/cases (18653) with CPT around -20 and -17 degrees Celsius for 18653 fridges/cases during 24 hours.

Figure 31 shows the CPT evolution for 1 month (31 days) for 68721 cases both LT and HT cases: the HT cases are the ones visible in the figure. The 1 month lasts from the date of 2017-01-05 and time 13:00:00 to the date 2017-02-06 and time 12:00:00. The data used in Figure 31 is taken from the IMS Evolve web system (Milton Keynes, UK) by hour (i.e. size of data 4 GB), and for comparison purposes minute by minute data would have been much bigger and harder to process (~240 GB). Figure 32 shows again the CPT data for 12 days for the full estate of 68721 cases/cases. It can be seen again the wave shapes of the 24 hour profiles visible also in Figure 31, which are because of the defrosts: the CPTs (i.e. around -17.5 degrees Celsius) for the LT cases are not visible as they are covered by the CPTs of the HT cases (i.e. around 5.5 degrees Celsius).

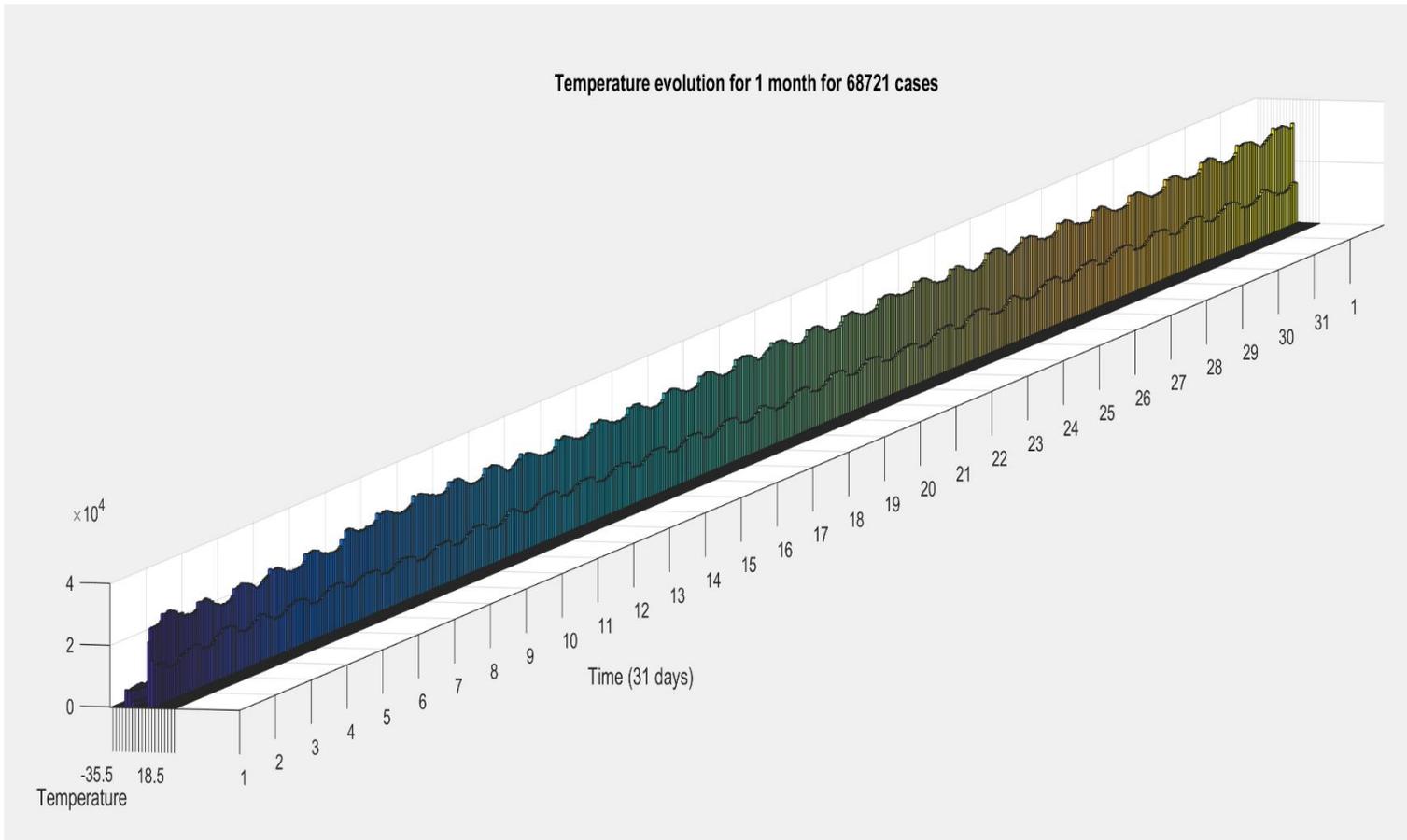

Figure 31. CPT temperature evolution for 1 month (31 days) for 68721 cases both LT and HT cases.

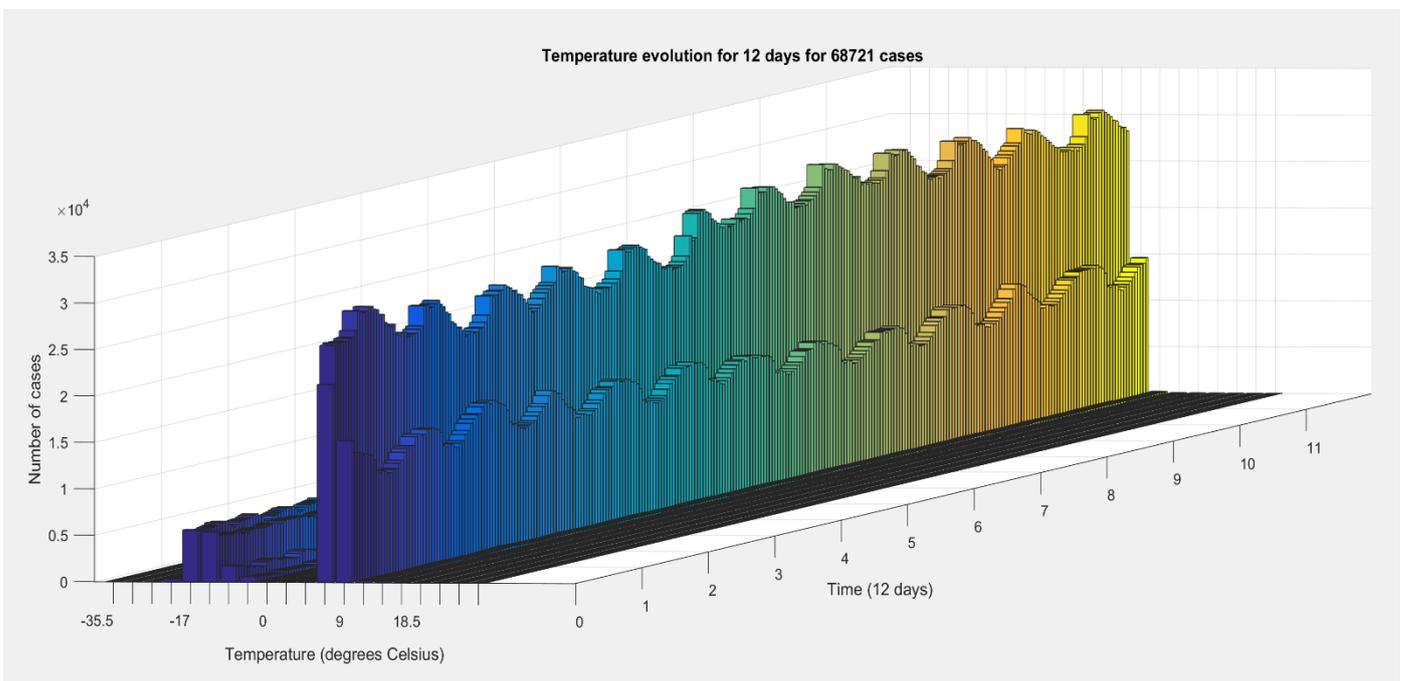

Figure 32. CPT temperature evolution for 12 days for 68721 cases both LT and HT cases.

*E. Seasonal fluctuations of warm-up temperature rates and defrosts time durtation for multiple HT and LT cases in multiple shops*

In the previous section, there were calculated the mean and the standard deviations of the warm-up rates (degrees Celsius per minute) of the HT and LT cases in various shops. Now there are added the numerical figures corresponding to the seasonal fluctuations between the warm season (June, July, August) and the cold season (December, January, February).

For a first shop, there can be noticed some differences for the means of the warm-up rates between the warm season and the cold season especially for the HT cases/cases. Per total over all the HT cases, there was a cumulated difference (i.e. sum up over all HT cases) in favor of the warm season (i.e. June, July, August) of 0.2235 degrees Celsius per minute for the mean (0.0045 degrees Celsius per minute per HT case) and 0.1017 degrees Celsius per minute for the standard deviation (Table 8).

| Case | Overall Yearly Mean | Standard deviations | Seasonal fluctuations (June, July, August) | | Seasonal fluctuations (December, January, February) | |
|---|---|---|---|---|---|---|
| | | | Mean | Standard deviations | Mean | Standard deviations |
| 1 | 0.0411 | 0.0234 | **0.0458** | 0.0253 | **0.0294** | 0.0139 |
| 2 | 0.0340 | 0.0187 | **0.0311** | 0.0163 | **0.0246** | 0.0124 |
| 3 | 0.0346 | 0.0165 | **0.0374** | 0.0166 | **0.0258** | 0.0098 |
| 4* | 0.0317 | 0.0154 | **0.0332** | 0.0147 | **0.0275** | 0.0166 |
| 5** | 0.0340 | 0.0166 | **0.0357** | 0.0181 | **0.0292** | 0.0138 |
| 6 | 0.0324 | 0.0102 | **0.0364** | 0.0133 | **0.0301** | 0.0065 |
| 7 | 0.0281 | 0.0048 | **0.0291** | 0.0049 | **0.0272** | 0.0047 |
| 8 | 0.0321 | 0.0052 | **0.0331** | 0.0051 | **0.0308** | 0.0045 |
| 9 | 0.0394 | 0.0050 | **0.0415** | 0.0050 | **0.0380** | 0.0043 |
| 10 | 0.0416 | 0.0058 | **0.0456** | 0.0053 | **0.0377** | 0.0036 |
| 11 | 0.0297 | 0.0056 | **0.0308** | 0.0050 | **0.0268** | 0.0045 |
| 12 | 0.0559 | 0.0091 | **0.0553** | 0.0088 | **0.0526** | 0.0078 |
| 13 | 0.0282 | 0.0168 | **0.0320** | 0.0201 | **0.0224** | 0.0121 |
| 14 | 0.0317 | 0.0063 | **0.0355** | 0.0060 | **0.0288** | 0.0035 |
| 15 | 0.0326 | 0.0066 | **0.0360** | 0.0059 | **0.0305** | 0.0060 |
| 16 | 0.0301 | 0.0064 | **0.0331** | 0.0063 | **0.0273** | 0.0048 |
| 17 | 0.0389 | 0.0064 | **0.0426** | 0.0062 | **0.0367** | 0.0037 |
| 18 | 0.0332 | 0.0064 | **0.0357** | 0.0067 | **0.0302** | 0.0051 |
| 19 | 0.0311 | 0.0054 | **0.0308** | 0.0051 | **0.0302** | 0.0042 |
| 20 | 0.0382 | 0.0067 | **0.0400** | 0.0063 | **0.0349** | 0.0048 |
| 21 | 0.0298 | 0.0056 | **0.0318** | 0.0058 | **0.0265** | 0.0038 |
| 22 | 0.0372 | 0.0054 | **0.0393** | 0.0057 | **0.0360** | 0.0037 |
| 23 | 0.0321 | 0.0054 | **0.0320** | 0.0059 | **0.0312** | 0.0044 |
| 24 | 0.0368 | 0.0072 | **0.0412** | 0.0056 | **0.0331** | 0.0050 |
| 25 | 0.0331 | 0.0067 | **0.0368** | 0.0063 | **0.0295** | 0.0044 |
| 26 | 0.0294 | 0.0056 | **0.0293** | 0.0061 | **0.0285** | 0.0043 |
| 27 | 0.0312 | 0.0071 | **0.0333** | 0.0076 | **0.0267** | 0.0045 |
| 28 | 0.0284 | 0.0065 | **0.0321** | 0.0059 | **0.0252** | 0.0050 |
| 29 | 0.0266 | 0.0063 | **0.0232** | 0.0059 | **0.0266** | 0.0054 |
| 30 | 0.0323 | 0.0054 | **0.0340** | 0.0055 | **0.0303** | 0.0045 |
| 31 | 0.0313 | 0.0053 | **0.0322** | 0.0061 | **0.0300** | 0.0039 |
| 32 | 0.0221 | 0.0063 | **0.0253** | 0.0056 | **0.0177** | 0.0042 |
| 33 | 0.0285 | 0.0052 | **0.0303** | 0.0055 | **0.0269** | 0.0039 |
| 34 | 0.0369 | 0.0050 | **0.0376** | 0.0058 | **0.0369** | 0.0042 |
| 35 | 0.0302 | 0.0058 | **0.0292** | 0.0070 | **0.0326** | 0.0046 |
| 36 | 0.0255 | 0.0063 | **0.0279** | 0.0062 | **0.0219** | 0.0045 |
| 37** | 0.0450 | 0.0309 | | | | |
| 38 | 0.0337 | 0.0049 | **0.0347** | 0.0055 | **0.0327** | 0.0036 |
| 39 | 0.0325 | 0.0050 | **0.0327** | 0.0053 | **0.0316** | 0.0031 |
| 40 | 0.0451 | 0.0061 | **0.0473** | 0.0062 | **0.0429** | 0.0043 |
| 41 | 0.0373 | 0.0057 | **0.0400** | 0.0055 | **0.0354** | 0.0037 |
| 42 | 0.0447 | 0.0058 | **0.0464** | 0.0043 | **0.0425** | 0.0048 |
| 43 | 0.0395 | 0.0058 | **0.0404** | 0.0057 | **0.0375** | 0.0036 |
| 44 | 0.0405 | 0.0058 | **0.0405** | 0.0060 | **0.0402** | 0.0041 |
| 45 | 0.0415 | 0.0053 | **0.0410** | 0.0048 | **0.0405** | 0.0044 |
| 46 | 0.0317 | 0.0051 | **0.0339** | 0.0052 | **0.0297** | 0.0039 |
| 47** | 0.0689 | 0.0256 | **0.0749** | 0.0313 | **0.0679** | 0.0274 |
| 48 | 0.0309 | 0.0071 | **0.0348** | 0.0061 | **0.0277** | 0.0060 |
| 49 | 0.0294 | 0.0057 | **0.0304** | 0.0053 | **0.0286** | 0.0050 |
| 50 | 0.0307 | 0.0065 | **0.0323** | 0.0062 | **0.0298** | 0.0053 |
| 51 | 0.0312 | 0.0062 | **0.0334** | 0.0067 | **0.0281** | 0.0038 |
| 52** | | | | | | |

Table 8. Warm-up mean and standard deviations of HT cases (shop 1) (degrees Celsius per minute).

There not observed big differences for the LT cases: over all the LT cases, the cumulated difference (i.e. sum up over all the LT cases) for the mean value is of only 0.0276 degrees Celsius per minute (0.0015 degrees Celsius per minute per LT case) and for the standard deviations of only 0.0030 degrees Celsius per minute (Table 9).

| Case | Overall Yearly Mean | Standard deviation | Seasonal fluctuations (June, July, August) | | Seasonal fluctuations (December, January, February) | |
|---|---|---|---|---|---|---|
| | | | Mean | Standard deviations | Mean | Standard deviations |
| 1 | 0.1163 | 0.0166 | 0.1088 | 0.0131 | 0.1120 | 0.0130 |
| 2 | 0.1027 | 0.0340 | 0.1044 | 0.0217 | 0.0789 | 0.0141 |
| 3 | 0.1771 | 0.0199 | 0.1708 | 0.0158 | 0.1679 | 0.0137 |
| 4 | 0.1450 | 0.0187 | 0.1400 | 0.0167 | 0.1592 | 0.0154 |
| 5 | 0.1604 | 0.0121 | 0.1611 | 0.0096 | 0.1548 | 0.0114 |
| 6 | 0.1591 | 0.0143 | 0.1577 | 0.0083 | 0.1540 | 0.0157 |
| 7 | 0.1812 | 0.0215 | 0.1833 | 0.0134 | 0.1851 | 0.0110 |
| 8 | 0.1910 | 0.0135 | 0.1849 | 0.0126 | 0.1899 | 0.0082 |
| 9 | 0.1392 | 0.0202 | 0.1312 | 0.0105 | 0.1277 | 0.0117 |
| 10 | 0.2056 | 0.0170 | 0.2044 | 0.0146 | 0.2002 | 0.0168 |
| 11 | 0.1512 | 0.0222 | 0.1394 | 0.0132 | 0.1484 | 0.0203 |
| 12 | 0.1559 | 0.0199 | 0.1462 | 0.0113 | 0.1649 | 0.0190 |
| 13 | 0.1310 | 0.0233 | 0.1136 | 0.0100 | 0.1322 | 0.0148 |
| 14 | 0.1319 | 0.0196 | 0.1279 | 0.0137 | 0.1397 | 0.0183 |
| 15 | 0.1436 | 0.0222 | 0.1357 | 0.0179 | 0.1615 | 0.0098 |
| 16 | 0.1460 | 0.0173 | 0.1438 | 0.0169 | 0.1397 | 0.0181 |
| 17 | 0.1387 | 0.0292 | 0.1467 | 0.0172 | 0.1094 | 0.0077 |
| 18 | 0.1598 | 0.0148 | 0.1556 | 0.0100 | 0.1576 | 0.0105 |

Table 9. Warm-up Mean and standard deviations of LT cases (shop 1) (degrees Celsius per minute).

In conclusion, differences could be noticed for the warm-up rates for the HT cases between the warm season (June, July, August) and the cold season (December, January, February).

In terms of duration of defrosts between the warm season (June, July, August) and the cold season (December, January, February), it is not observed a significant difference (i.e. only 2.4 minutes cumulated over all the LT cases) for the mean values of defrosts for the LT cases (Table 10). This might be explained mainly by the fact that the LT cases have doors so the ambient temperature influences less the CPT for refrigeration cases, which have with doors. There is only 0.13 minutes per LT case more time to defrost in the warm season than in the cold season that is because probably people are using less the LT cases in the summer season and are buying more fresh food.

| Case | Seasonal fluctuations (June, July, August) | Seasonal fluctuations (December, January, February) |
|---|---|---|
| | Mean | Mean |
| 1 | 29.35 | 31.13 |
| 2 | 13.39 | 12.06 |
| 3 | 27.26 | 28.37 |
| 4 | 45.02 | 30.13 |
| 5 | 41.58 | 42.82 |
| 6 | 37.66 | 37.47 |
| 7 | 32.75 | 31.90 |
| 8 | 27.70 | 30.26 |
| 9 | 29.24 | 27.22 |
| 10 | 32.53 | 29.93 |
| 11 | 34.30 | 39.26 |
| 12 | 28.80 | 29.41 |
| 13 | 29.40 | 28.60 |
| 14 | 30.76 | 28.28 |
| 15 | 30.49 | 24.73 |
| 16 | 30.47 | 31.60 |
| 17 | 43.72 | 58.49 |
| 18 | 30.06 | 30.42 |

Table 10. Time duration of defrost LT cases – mean values (shop 1).

In terms of duration of defrosts between the cold season (December, January, February) and the warm season (June, July, August) for the HT cases (Table 11), there seems to be a higher difference consisting of 22.94 minutes (i.e. summed up over all the HT cases) for the means of the duration of defrosts. This corresponds to 0.44 minutes per HT case more time needed to defrost in the cold season than in the warm season. This means that it is needed (i.e. 0.44 minutes on average) more time to defrost in the cold season than in the warm season for the HT cases.

| Case | Seasonal fluctuations (June, July, August) | Seasonal fluctuations (December, January, February) |
|---|---|---|
| | Mean | Mean |
| 1 | 5.39 | 6.01 |
| 2 | 14.80 | 15.14 |
| 3 | 4.44 | 4.68 |
| 4** | 4.43 | 4.63 |
| 5** | 3.91 | 4.42 |
| 6 | 22.37 | 28.37 |
| 7 | 28.69 | 28.88 |
| 8 | 38.41 | 38.76 |
| 9 | 43.38 | 43.75 |
| 10 | 42.02 | 43.44 |
| 11 | 43.64 | 43.62 |
| 12 | 26.23 | 25.27 |
| 13 | 4.74 | 6.44 |
| 14 | 43.07 | 42.57 |
| 15 | 28.59 | 33.76 |
| 16 | 28.50 | 28.88 |
| 17 | 43.11 | 43.71 |
| 18 | 28.58 | 28.91 |
| 19 | 28.44 | 28.90 |
| 20 | 38.45 | 38.43 |
| 21 | 33.16 | 33.57 |
| 22 | 43.21 | 43.62 |
| 23 | 28.52 | 28.84 |
| 24 | 43.32 | 28.79 |
| 25 | 28.37 | 28.65 |
| 26 | 28.45 | 28.73 |
| 27 | 28.46 | 28.74 |
| 28 | 28.47 | 28.66 |
| 29 | 28.36 | 28.62 |
| 30 | 33.50 | 33.68 |
| 31 | 43.23 | 43.58 |
| 32 | 28.41 | 28.53 |
| 33 | 28.49 | 28.67 |
| 34 | 41.85 | 43.62 |
| 35 | 28.43 | 28.69 |
| 36 | 28.32 | 28.52 |
| 37** | | |
| 38 | 42.57 | 43.59 |
| 39 | 42.85 | 43.75 |
| 40 | 40.89 | 43.37 |
| 41 | 43.25 | 43.50 |
| 42 | 34.91 | 41.51 |
| 43 | 43.04 | 43.56 |
| 44 | 42.82 | 43.44 |
| 45 | 42.17 | 43.40 |
| 46 | 42.47 | 43.51 |
| 47** | 2.51 | 3.16 |
| 48 | 28.29 | 28.68 |
| 49 | 43.04 | 43.00 |
| 50 | 42.91 | 43.30 |
| 51 | 42.91 | 43.48 |
| 52** | 29.24 | 27.22 |

Table 11. Time duration of defrost of HT cases – mean value (shop 1).

A second shop, which has 115 refrigeration cases/cases is processed. The data covers 1 year. There are 31 LT cases and 84 HT cases. There is a mean for the warm-up rates for LT cases of 0.1526 and a standard deviation of 0.0156 degrees Celsius per minute (Table 12).

| Case | Mean | Standard deviation | Seasonal fluctuations (June, July, August) | | Seasonal fluctuations (December, January, February) | |
|---|---|---|---|---|---|---|
| | | | Mean | Standard deviations | Mean | Standard deviations |
| 1 | 0.1336 | 0.0087 | 0.1394 | 0.0071 | 0.1297 | 0.0066 |
| 2 | 0.1152 | 0.0072 | 0.1116 | 0.0057 | 0.1198 | 0.0054 |
| 3 | 0.1348 | 0.0165 | 0.1275 | 0.0084 | 0.1433 | 0.0181 |
| 4 | 0.1771 | 0.0274 | 0.1737 | 0.0079 | 0.1817 | 0.0510 |
| 5 | 0.1248 | 0.0175 | 0.1197 | 0.0131 | 0.1232 | 0.0202 |
| 6 | 0.1525 | 0.0199 | 0.1456 | 0.0199 | 0.1573 | 0.0187 |
| 7 | 0.2034 | 0.0113 | 0.2032 | 0.0088 | 0.2019 | 0.0163 |
| 8 | 0.1396 | 0.0099 | 0.1304 | 0.0060 | 0.1465 | 0.0097 |
| 9 | 0.1387 | 0.0174 | 0.1223 | 0.0063 | 0.1538 | 0.0118 |
| 10 | 0.1234 | 0.0140 | 0.1319 | 0.0123 | 0.1229 | 0.0109 |
| 11 | 0.1416 | 0.0103 | 0.1463 | 0.0051 | 0.1421 | 0.0079 |
| 12 | 0.1500 | 0.0228 | 0.1508 | 0.0069 | 0.1585 | 0.0156 |
| 13 | 0.1237 | 0.0161 | 0.1204 | 0.0085 | 0.1355 | 0.0153 |
| 14 | 0.1175 | 0.0099 | 0.1136 | 0.00787 | 0.1248 | 0.0068 |
| 15 | 0.0961 | 0.0131 | 0.0971 | 0.0103 | 0.0930 | 0.0130 |
| 16 | 0.2185 | 0.0144 | 0.2222 | 0.0141 | 0.2176 | 0.0089 |
| 17 | 0.2174 | 0.0152 | 0.2228 | 0.0141 | 0.2129 | 0.0141 |
| 18 | 0.1354 | 0.0141 | 0.1422 | 0.0123 | 0.1368 | 0.0194 |

| 19 | 0.1418 | 0.0127 | 0.1418 | 0.0167 | 0.1403 | 0.0088 |
|----|--------|--------|--------|--------|--------|--------|
| 20 | 0.1362 | 0.0137 | 0.1422 | 0.0124 | 0.1382 | 0.0146 |
| 21 | 0.1817 | 0.0212 | 0.1865 | 0.0153 | 0.1850 | 0.0242 |
| 22 | 0.1626 | 0.0143 | 0.1656 | 0.0150 | 0.1582 | 0.0156 |
| 23 | 0.1163 | 0.0191 | 0.1073 | 0.0236 | 0.1215 | 0.0092 |
| 24 | 0.1897 | 0.0160 | 0.1861 | 0.0141 | 0.1940 | 0.0139 |
| 25 | 0.1425 | 0.0191 | 0.1496 | 0.0216 | 0.1375 | 0.0163 |
| 26 | 0.1188 | 0.0108 | 0.1222 | 0.0114 | 0.1170 | 0.0104 |
| 27 | 0.1751 | 0.0156 | 0.1750 | 0.0119 | 0.1765 | 0.0213 |
| 28 | 0.1284 | 0.0117 | 0.1334 | 0.0105 | 0.1222 | 0.0113 |
| 29 | 0.2037 | 0.0217 | 0.1976 | 0.0127 | 0.2078 | 0.0281 |
| 30 | 0.2078 | 0.0158 | 0.2052 | 0.0108 | 0.2101 | 0.0177 |
| 31 | 0.1819 | 0.0253 | 0.1638 | 0.0114 | 0.1824 | 0.0278 |

Table 12. Warm-up Mean and standard deviations of LT cases including seasonal fluctuations (shop 2) (degrees Celsius per minute).

There is also a mean for the warm-up rates for HT (Table 13) cases of 0.0364 and a standard deviation of 0.0080 degrees Celsius per minute. The HT cases with a double star do not have defrost status available in the dataset. There can be noted some differences for the means of the warm-up rates between the warm season and the cold season for the HT cases/cases: per total over all the HT cases, there was a cumulated difference (i.e. summed up over all HT cases) in favour of the warm season (i.e. June, July, August) of 0.4247 degrees Celsius per minute for the mean value of the warm-up rate.

| Case | Mean | Standard deviation | Seasonal fluctuations (June, July, August) | | Seasonal fluctuations (December, January, February) | |
|------|------|-------------------|------|------|------|------|
| | | | Mean | Standard deviations | Mean | Standard deviations |
| 1 | 0.0309 | 0.0052 | 0.0350 | 0.0038 | 0.0281 | 0.0046 |
| 2 | 0.0306 | 0.0048 | 0.0343 | 0.0039 | 0.0277 | 0.0030 |
| 3 | 0.0302 | 0.0043 | 0.0327 | 0.0041 | 0.0285 | 0.0035 |
| 4 | 0.0243 | 0.0042 | 0.0247 | 0.0043 | 0.0235 | 0.0039 |
| 5 | 0.0339 | 0.0054 | 0.0359 | 0.0047 | 0.0327 | 0.0068 |
| 6 | 0.0353 | 0.0063 | 0.0402 | 0.0060 | 0.0319 | 0.0043 |
| 7 | 0.0323 | 0.0066 | 0.0370 | 0.0066 | 0.0285 | 0.0041 |
| 8 | 0.0290 | 0.0056 | 0.0331 | 0.0052 | 0.0255 | 0.0038 |
| 9 | 0.0402 | 0.0107 | 0.0409 | 0.0141 | 0.0395 | 0.0064 |
| 10 | 0.0313 | 0.0059 | 0.0338 | 0.0078 | 0.0309 | 0.0033 |
| 11 | 0.0406 | 0.0079 | 0.0458 | 0.0064 | 0.0359 | 0.0078 |
| 12 | 0.0368 | 0.0114 | 0.0419 | 0.0052 | 0.0231 | 0.0123 |
| 13 | 0.0305 | 0.0050 | 0.0323 | 0.0044 | 0.0282 | 0.0060 |
| 14 | 0.0413 | 0.0069 | 0.0435 | 0.0065 | 0.0406 | 0.0067 |
| 15 | 0.0312 | 0.0061 | 0.0318 | 0.0065 | 0.0315 | 0.0052 |
| 16 | 0.0331 | 0.0067 | 0.0341 | 0.0083 | 0.0333 | 0.0062 |
| 17 | 0.0445 | 0.0084 | 0.0487 | 0.0082 | 0.0415 | 0.0065 |
| 18 | 0.0412 | 0.0108 | 0.0427 | 0.0128 | 0.0408 | 0.0078 |
| 19 | 0.0392 | 0.0071 | 0.0402 | 0.0065 | 0.0393 | 0.0066 |
| 20 | 0.0348 | 0.0064 | 0.0386 | 0.0065 | 0.0308 | 0.0048 |
| 21 | 0.0348 | 0.0068 | 0.0367 | 0.0079 | 0.0335 | 0.0075 |
| 22 | 0.0317 | 0.0036 | 0.0316 | 0.0036 | 0.0330 | 0.0039 |
| 23 | 0.0453 | 0.0053 | 0.0480 | 0.0049 | 0.0429 | 0.0057 |
| 24 | 0.0259 | 0.0058 | 0.0307 | 0.0047 | 0.0228 | 0.0050 |
| 25 | 0.0285 | 0.0056 | 0.0327 | 0.0055 | 0.0257 | 0.0035 |
| 26 | 0.0390 | 0.0079 | 0.0455 | 0.0052 | 0.0332 | 0.0076 |
| 27 | 0.0305 | 0.0058 | 0.0342 | 0.0054 | 0.0273 | 0.0037 |
| 28 | 0.0305 | 0.0054 | 0.0323 | 0.0057 | 0.0296 | 0.0048 |
| 29 | 0.0325 | 0.0048 | 0.0348 | 0.0049 | 0.0305 | 0.0040 |
| 30 | 0.0293 | 0.0059 | 0.0337 | 0.0047 | 0.0263 | 0.0060 |
| 31 | 0.0369 | 0.0048 | 0.0398 | 0.0051 | 0.0342 | 0.0037 |
| 32 | 0.0405 | 0.0066 | 0.0461 | 0.0056 | 0.0352 | 0.0044 |
| 33 | 0.0493 | 0.0072 | 0.0533 | 0.0075 | 0.0454 | 0.0067 |
| 34 | 0.0483 | 0.0081 | 0.0501 | 0.0084 | 0.0443 | 0.0067 |
| 35 | 0.0442 | 0.0078 | 0.0490 | 0.0078 | 0.0407 | 0.0063 |
| 36 | 0.0482 | 0.0069 | 0.0494 | 0.0062 | 0.0490 | 0.0083 |
| 37 | 0.0353 | 0.0072 | 0.0344 | 0.0073 | 0.0368 | 0.0069 |
| 38 | 0.0655 | 0.0122 | 0.0655 | 0.0122 | 0.0653 | 0.0141 |
| 39** | | | | | | |
| 40 | 0.0278 | 0.0050 | 0.0299 | 0.0054 | 0.0266 | 0.0040 |
| 41 | 0.0900 | 0.0496 | 0.1271 | 0.0496 | 0.0638 | 0.0284 |
| 42 | 0.0879 | 0.0178 | 0.0874 | 0.0138 | 0.0908 | 0.0228 |
| 43 | 0.0988 | 0.0273 | 0.0990 | 0.0264 | 0.0987 | 0.0286 |
| 44 | 0.0842 | 0.0211 | 0.0798 | 0.0194 | 0.0867 | 0.0222 |
| 45** | | | | | | |
| 46 | 0.0399 | 0.0246 | 0.0403 | 0.0286 | 0.0386 | 0.0193 |
| 47 | 0.0528 | 0.0331 | 0.0561 | 0.0364 | 0.0458 | 0.0258 |
| 48 | 0.0590 | 0.0189 | 0.0550 | 0.0214 | 0.0604 | 0.0161 |
| 49 | 0.0424 | 0.0057 | 0.0442 | 0.0058 | 0.0411 | 0.0054 |
| 50 | 0.0327 | 0.0064 | 0.0352 | 0.0051 | 0.0277 | 0.0062 |
| 51 | 0.0309 | 0.0062 | 0.0359 | 0.0047 | 0.0302 | 0.0084 |
| 52 | 0.0197 | 0.0053 | 0.0230 | 0.0050 | 0.0187 | 0.0036 |
| 53 | 0.0321 | 0.0045 | 0.0342 | 0.0048 | 0.0308 | 0.0035 |
| 54 | 0.0230 | 0.0061 | 0.0256 | 0.0062 | 0.0226 | 0.0055 |
| 55 | 0.0312 | 0.0077 | 0.0337 | 0.0066 | 0.0297 | 0.0104 |
| 56 | 0.0470 | 0.0063 | 0.0509 | 0.0062 | 0.0431 | 0.0045 |
| 57 | 0.0362 | 0.0062 | 0.0402 | 0.0061 | 0.0336 | 0.0042 |
| 58 | 0.0279 | 0.0063 | 0.0338 | 0.0051 | 0.0236 | 0.0040 |
| 59 | 0.0143 | 0.0043 | 0.0140 | 0.0049 | 0.0135 | 0.0038 |
| 60 | 0.0071 | 0.0051 | 0.0080 | 0.0049 | 0.0051 | 0.0025 |
| 61 | 0.0205 | 0.0060 | 0.0216 | 0.0064 | 0.0209 | 0.0053 |
| 62 | 0.0367 | 0.0063 | 0.0398 | 0.0062 | 0.0353 | 0.0050 |
| 63 | 0.0252 | 0.0093 | 0.0345 | 0.0063 | 0.0184 | 0.0063 |
| 64 | 0.0266 | 0.0055 | 0.0302 | 0.0061 | 0.0241 | 0.0035 |
| 65 | 0.0368 | 0.0137 | 0.0384 | 0.0140 | 0.0363 | 0.0140 |

| 66 | 0.0279 | 0.0056 | 0.0268 | 0.0050 | 0.0287 | 0.0061 |
| --- | --- | --- | --- | --- | --- | --- |
| 67 | 0.0244 | 0.0065 | 0.0219 | 0.0066 | 0.0254 | 0.0057 |
| 68 | 0.0324 | 0.0053 | 0.0345 | 0.0053 | 0.0307 | 0.0045 |
| 69 | 0.0533 | 0.0082 | 0.0571 | 0.0070 | 0.0494 | 0.0068 |
| 70 | 0.0235 | 0.0043 | 0.0261 | 0.0045 | 0.0217 | 0.0034 |
| 71 | 0.0240 | 0.0038 | 0.0269 | 0.0033 | 0.0220 | 0.0025 |
| 72 | 0.0259 | 0.0037 | 0.0282 | 0.0036 | 0.0240 | 0.0028 |
| 73 | 0.0290 | 0.0041 | 0.0314 | 0.0039 | 0.0276 | 0.0033 |
| 74 | 0.0223 | 0.0051 | 0.0237 | 0.0051 | 0.0237 | 0.0036 |
| 75 | 0.0200 | 0.0033 | 0.0220 | 0.0032 | 0.0186 | 0.0025 |
| 76 | 0.0299 | 0.0038 | 0.0251 | 0.0040 | 0.0213 | 0.0028 |
| 77 | 0.0201 | 0.0042 | 0.0234 | 0.0035 | 0.0174 | 0.0033 |
| 78 | 0.0181 | 0.0036 | 0.0204 | 0.0035 | 0.0167 | 0.0032 |
| 79 | 0.0419 | 0.0059 | 0.0464 | 0.0052 | 0.0380 | 0.0039 |
| 80 | 0.0450 | 0.0076 | 0.0490 | 0.0068 | 0.0423 | 0.0081 |
| 81 | 0.0302 | 0.0053 | 0.0308 | 0.0053 | 0.0288 | 0.0051 |
| 82 | 0.0286 | 0.0062 | 0.0313 | 0.0050 | 0.0267 | 0.0080 |
| 83 | 0.0309 | 0.0053 | 0.0346 | 0.0051 | 0.0284 | 0.0036 |
| 84 | 0.0392 | 0.0061 | 0.0435 | 0.0053 | 0.0362 | 0.0048 |

Table 13. Warm-up Mean and standard deviations of HT cases including seasonal fluctuations (shop 2) (degrees Celsius per minute).

Smaller differences could be noticed for the LT cases (Table 14): over all the LT cases, the cumulated difference (i.e. sum up over all the LT cases) for the means is of only 0.09 degrees Celsius per minute. In terms of duration of defrosts (Table 14) between the warm season (June, July, August) and the cold season (December, January, February), there is not a very big difference, that is only 18 minutes cumulated over all the LT cases. This is 0.58 minutes per case on average in comparison between the warm season and the cold season (i.e. it is needed more time to defrost in the warm season than in the cold season) for the mean values of defrosts for the LT cases. This might be explained again mainly by the fact that the LT cases have doors so the ambient temperature influences less the CPTs for cases with doors. Finally, as there is slightly more time needed to defrost in the warm season for the LT cases (i.e. 0.58 minute per case), this might suggest again that the LT cases are used less in the warm season as the people buy more fresh food in the warm season.

| Case | Seasonal fluctuations (June, July, August) | Seasonal fluctuations (December, January, February) |
| --- | --- | --- |
| | Mean | Mean |
| 1 | 68.26 | 68.23 |
| 2 | 68.65 | 68.74 |
| 3 | 68.08 | 66.95 |
| 4 | 43.22 | 42.11 |
| 5 | 38.30 | 38.06 |
| 6 | 28.30 | 28.12 |
| 7 | 42.45 | 40.75 |
| 8 | 43.31 | 42.82 |
| 9 | 43.51 | 43.12 |
| 10 | 43.36 | 43.09 |
| 11 | 43.30 | 42.96 |
| 12 | 43.35 | 35 |
| 13 | 43.39 | 42.97 |
| 14 | 43.24 | 43.11 |
| 15 | 38.09 | 37.80 |
| 16 | 24.05 | 22.05 |
| 17 | 24.00 | 22.59 |
| 18 | 29.93 | 26.72 |
| 19 | 27.19 | 26.94 |
| 20 | 29.96 | 35.28 |
| 21 | 26.45 | 33.10 |
| 22 | 24.16 | 23.96 |
| 23 | 24.77 | 22.86 |
| 24 | 24.86 | 22.65 |
| 25 | 22.94 | 21.95 |
| 26 | 22.17 | 21.69 |
| 27 | 24.96 | 24.90 |
| 28 | 22.26 | 21.28 |
| 29 | 30.34 | 28.14 |
| 30 | 26.15 | 25.67 |
| 31 | 23.38 | 24.62 |

Table 14. Time duration of defrost for LT cases – mean values (minutes) (shop 2).

In terms of duration of defrosts between the cold season (December, January, February) and the warm season (June, July, August) for the HT cases (Table 15), there seems to be a higher difference of 74 minutes for the means of the duration of defrosts cumulated over all the HT cases. This is 0.91 minutes per case on average between the cold season and the warm season (i.e. it is needed less time in the warm season to defrost). This may mean that it is needed more time to defrost in the cold season than in the warm season for the HT cases.

| Case | Seasonal fluctuations (June, July, August) | Seasonal fluctuations (December, January, February) |
|---|---|---|
| | Mean | Mean |
| 1 | 43.29 | 42.85 |
| 2 | 43.41 | 43.07 |
| 3 | 43.24 | 43.04 |
| 4 | 43.36 | 43.10 |
| 5 | 43.31 | 42.65 |
| 6 | 43.26 | 43.18 |
| 7 | 43.31 | 43.13 |
| 8 | 43.37 | 43.08 |
| 9 | 10.56 | 21.68 |
| 10 | 24.06 | 41.94 |
| 11 | 28.27 | 28.15 |
| 12 | 25.39 | 28.19 |
| 13 | 28.46 | 28.20 |
| 14 | 24.67 | 28.22 |
| 15 | 21.94 | 27.60 |
| 16 | 14.18 | 23.57 |
| 17 | 27.24 | 28.16 |
| 18 | 15.58 | 23.30 |
| 19 | 20.79 | 24.62 |
| 20 | 25.33 | 28.13 |
| 21 | 26.57 | 39.72 |
| 22 | 43.37 | 43.25 |
| 23 | 41.56 | 43.03 |
| 24 | 43.19 | 42.76 |
| 25 | 43.42 | 43.02 |
| 26 | 43.30 | 42.79 |
| 27 | 43.43 | 43.07 |
| 28 | 43.15 | 43.04 |
| 29 | 43.36 | 43.11 |
| 30 | 43.39 | 42.69 |
| 31 | 43.38 | 43.22 |
| 32 | 43.23 | 43.19 |
| 33 | 34.84 | 42.45 |
| 34 | 40.07 | 42.60 |
| 35 | 42.49 | 42.91 |
| 36 | 17 | 19.11 |
| 37 | 10.88 | 11.58 |
| 38 | 8.02 | 12.33 |
| 39** | | |
| 40 | 43.26 | 42.96 |
| 41 | 2.01 | 2.06 |
| 42 | 9.30 | 9.20 |
| 43 | 5.93 | 5.88 |
| 44 | 6.39 | 6.50 |
| 45** | | |
| 46 | 7.69 | 5.96 |
| 47 | 9.82 | 5.13 |
| 48 | 14.39 | 11.30 |
| 49 | 43.34 | 43.15 |
| 50 | 43.17 | 42.81 |
| 51 | 43.29 | 42.71 |
| 52 | 43.43 | 43.12 |

| 53 | 43.24 | 42.95 |
|---|---|---|
| 54 | 43.35 | 43.15 |
| 55 | 43.40 | 42.67 |
| 56 | 43.37 | 43.26 |
| 57 | 43.46 | 43.22 |
| 58 | 43.47 | 43.69 |
| 59 | 43.25 | 42.73 |
| 60 | 43.42 | 43.16 |
| 61 | 43.34 | 42.94 |
| 62 | 43.30 | 43.14 |
| 63 | 43.32 | 42.65 |
| 64 | 43.35 | 43.22 |
| 65 | 13.38 | 8.92 |
| 66 | 37.48 | 33.70 |
| 67 | 16.70 | 18.14 |
| 68 | 43.34 | 43.15 |
| 69 | 42.10 | 43.18 |
| 70 | 23.85 | 28.12 |
| 71 | 43.23 | 43.16 |
| 72 | 43.23 | 42.97 |
| 73 | 43.25 | 43.14 |
| 74 | 43.33 | 42.72 |
| 75 | 43.28 | 43.30 |
| 76 | 43.30 | 43.18 |
| 77 | 43.28 | 43.67 |
| 78 | 43.10 | 42.76 |
| 79 | 43.25 | 43.13 |
| 80 | 42.51 | 42.79 |
| 81 | 43.22 | 43.16 |
| 82 | 43.27 | 42.71 |
| 83 | 43.21 | 43.28 |
| 84 | 38.74 | 38.65 |

Table 15. Time duration of defrost for HT cases – mean values (minutes) (shop 2).

## III. CONTROLLING AND MONITORING SYSTEM FOR MASSIVE NETWORKS OF HT AND LT FOOD REFRIGERATION SYSTEMS

The controlling and monitoring of food chains on a national massive scale poses significant challenges, which must take into account the features and the differences between the LT and the HT cases when designing a decision system for monitoring and control of 100000s of refrigeration cases. Such a decision system requires considerable novel IT architecture development based on Big Data and IoTs to handle very large and rapid data streams together with the potential integration of cold energy storage systems, robust and effective control system architecture, novel metering, integration into the National Grid of a country, and a high level engineering standards to prevent failures all in a food safe environment.

In this context, Big Data [19-22] has four main characteristics which can be found also in the controlling and monitoring of food chains involving > 100000s of HT and LT refrigeration cases. In order to monitor these HT and LT cases, multiple sensors are to be mounted on the respective individual refrigeration cases in order to send data (e.g. pressure measurements, temperature measurements, alarms) though the IoTs. The volume of the data coming from the refrigeration cases can be really huge and something around 20-30 billion data points can be obtained per day with tens of thousands of concurrent system users (i.e. volume – first characteristic). This data can come in through at a very high band width private network and would have to be processed very fast within seconds so that first, to detect any anomalous events with regard to the refrigeration cases, and to enable extremely rapid remedial responses and second to also deal with an unstable electric grid (i.e. velocity). This would be the second characteristic of Big Data in this response situation.

Monitoring and control of this large system of HT and LT cases may not be a problem as the communications with this system can be made within 40ms for any refrigeration case located in any part of the country.

Moreover, there can be a wide variety of data because of the differences described in the previous chapter between the HT and the LT refrigeration cases. This variety of data would be represented by information about temperatures, pressures, type of refrigeration cases (e.g. for meat, for dairy products, frozen food, etc), power consumption, HVAC systems, lights and many other, such as alternative control strategies from differing fridge manufactures (i.e. the third characteristic - variety).

Finally, at some point during the functioning of the entire monitoring and decision system, there could be some data uncertainty for example from the malfunction of the mounted sensors or from problems with comunicating the data through the IoT (i.e. the fourth characteristic - veracity).

Alltogether, these four characteristics depict an overal situation where data processing paradigm can be used in order to alleviate any emergency situations that can appear in the monitoring and control of these large systems of HT and LT refrigeration cases.

Figure 33 shows a possible structure of a decision system for monitoring, controlling and supporting the maintenance of > 100,000 refrigeration HT and LT cases across an entire country. Each refrigeration case can be networked to a local, in store, data control centre through which each machine can be controlled remotely, together with recording of temperature data to ensure food safety in agreement with regulations, whilst supporting machine maintenance by performance and alarming systems monitoring. The control capability can operate over a temperature tolerance, commensurate with food safety, and therefore provide energy control across the entire range of refrigeration cases.

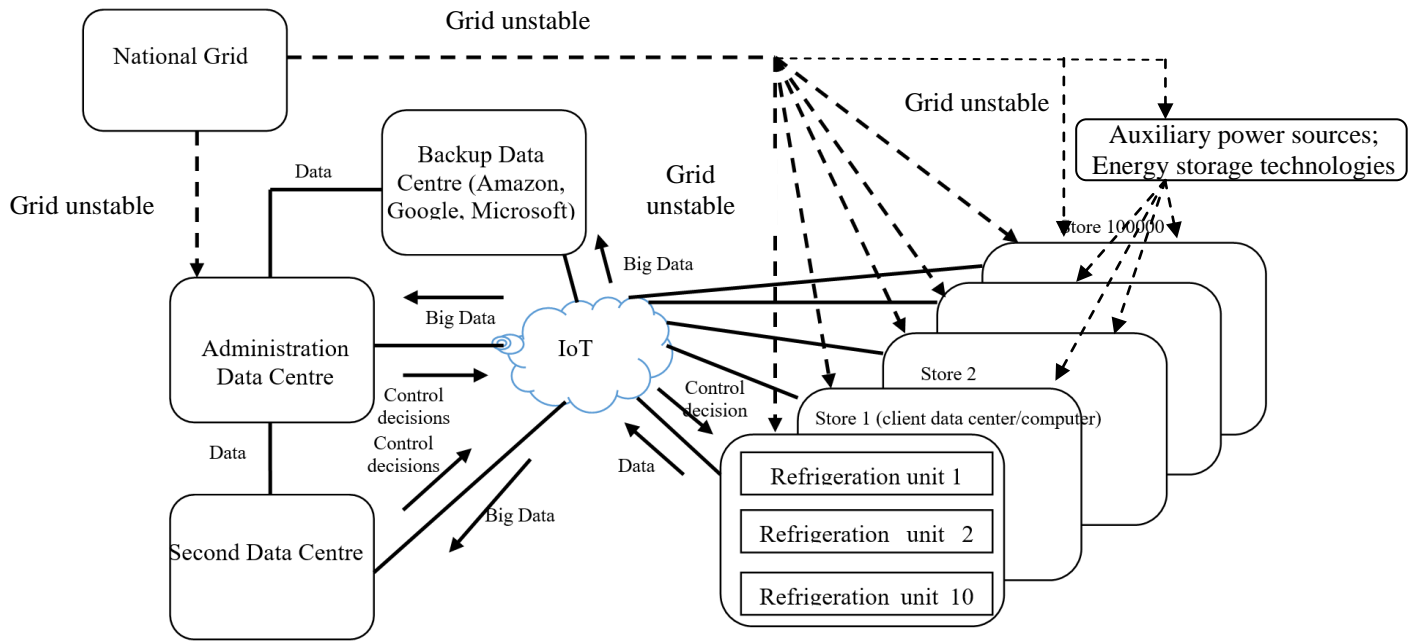

Figure 33. Decision system for monitoring and control of 100000s of HT and LT refrigeration cases involving Big Data and IoT.

The decision system includes a main Administration Data Center where the data can arrive in real time and some important decisions (i.e. switching on and off the refrigeration cases) can be taken with regard to the functioning of the HT and LT refrigeration cases including the management of the situations which arise from fluctuations in power demands and generations output from the National Grid and therefore resulting for example in excess or power loss for any of the food retailer shops and the HT/LT refrigeration cases. In addition to the Administration Data Center, a second Data Center where the data can be archived and from this second data centre, control decisions based on the longer term trend data can be taken, that is performance and alarm monitoring. The scope of this second Data Center is also to support the Administration Data Center especially in the diurnal periods. The second Data Center can back up the data to dedicated data centers from Amazon, Google or Microsoft, which provide more general cloud computing services.

Finally, control decisions can be taken at the local level of a store where a Client Data Centre exists. The scope of this Client Data centre is also to forward the data to the other data centres (e.g. Administrative Data Centre, Backup Data Centre).

In Figure 33 the direction of arrows shows the direction of the flow of the control decisions or the data flow such as the decisions taken from the Administration Data Centre are travelling down through a private secure network or IoT to the Client Data Centre associated to a store where the administrator of the shop can take the optimum decisions. Therefore, for the National Grid in the situation of an unstable electrical grid, this information will be forwarded within milliseconds both to the central Administration Data Centre and to the Client Data Centre Stores where the optimum decisions can be taken for the existent HT and LT refrigeration cases. Figure 33 includes also for example auxiliary power sources or energy storage technologies, which can be used in the critical situations when the grid may become unstable.

As a recent example, a higher level framework entitled "Nemesyst" [23] was proposed, which combines IoT, Big Data and deep learning [24-26] and which tries to optimise the functioning of individual machines in refrigeration massive networks consisting of different LT and HT cases.

## IV. Conclusions

This paper used data science to mainly investigate the temperature profiles of massive networks of food refrigeration systems, which exist in food retailer shops. The study showed important differences among the various types of refrigeration cases (HT cases versus LT cases) but also during the operation of these cases during various time periods (e.g. month, year, warm season versus winter season) such as higher warm-up rates in degrees Celsius per minute for the LT cases compared to the HT cases or more time required to defrost for the HT cases in the cold season than in the warm season. These differences must be taken into account when designing a decision system for monitoring and control of 100000s of refrigeration cases, and which to be based on Big Data and IoT.

Future work will involve further developing the decision system for monitoring and control of 100000s of HT and LT refrigeration cases involving Big Data and IoT.

Acknowledgment

This research was supported by Innovate UK (Grant No. 54043-400273). The author would like to thank Tesco, IMS Evolve and ECH Engineering.